\documentclass[iop]{emulateapj}

\usepackage{natbib,epsfig}
\usepackage{morefloats}
\shorttitle{Kepler data guide}

\begin{document}

\title{Demystifying Kepler Data: A Primer for Systematic Artifact Mitigation}

\author{K. Kinemuchi}
\affil{Bay Area Environmental Research Institute \& NASA-Ames Research Center}
\affil{Mail Stop 244-30, Moffett Field, CA 94035}
\email{karen.kinemuchi@nasa.gov}

\author{T. Barclay}
\affil{Bay Area Environmental Research Institute \& NASA-Ames Research Center}
\affil{Mail Stop 244-30, Moffett Field, CA 94035}
\email{thomas.barclay@nasa.gov}

\author{M. Fanelli}
\affil{Bay Area Environmental Research Institute \& NASA-Ames Research Center}
\affil{Mail Stop 245-1, Moffett Field, CA 94035}
\email{michael.n.fanelli@nasa.gov}

\author{J. Pepper}
\affil{Vanderbilt University}
\affil{Department of Physics and Astronomy, Nashville, TN 37235}
\email{joshua.pepper.v@gmail.com}

\author{M. Still}
\affil{Bay Area Environmental Research Institute \& NASA-Ames Research Center}
\affil{Mail Stop 244-30, Moffett Field, CA 94035}
\email{martin.still@nasa.gov}

\and

\author{Steve B. Howell}
\affil{NASA-Ames Research Center}
\affil{Mail Stop 244-30, Moffett Field, CA 94035}
\email{steve.b.howell@nasa.gov}

\begin{abstract}
The Kepler spacecraft has collected data of high photometric precision and cadence almost continuously since operations began on 2009 May 2. Primarily designed to detect planetary transits and asteroseismological signals from solar-like stars, Kepler has provided high quality data for many areas of investigation. Unconditioned simple aperture time-series photometry are however affected by systematic structure.  Examples of these systematics are differential velocity aberration, thermal gradients across the spacecraft, and pointing variations. While exhibiting some impact on Kepler's primary science, these systematics can critically handicap potentially ground-breaking scientific gains in other astrophysical areas, especially over long timescales greater than 10 days. As the data archive grows to provide light curves for $10^5$ stars of many years in length, Kepler will only fulfill its broad potential for stellar astrophysics if these systematics are understood and mitigated.  Post-launch developments in the Kepler archive, data reduction pipeline and open source data analysis software have occurred to remove or reduce systematic artifacts. This paper provides a conceptual primer for users of the Kepler data archive to understand and recognize systematic artifacts within light curves and some methods for their removal. Specific examples of artifact mitigation are provided using data available within the archive. Through the methods defined here, the Kepler community will find a road map to maximizing the quality and employment of the Kepler legacy archive.

\end{abstract}


\section{Introduction}
The Kepler spacecraft was launched on 2009 March 6 with a potential operational lifetime limited by its approximately 10 year supply of propellant (Koch et al. 2010). Kepler's primary objective is to determine the frequency of Earth-sized planets within the habitable zone of solar-like stars, achieved by detecting planetary transits of stars within high-precision time-series photometry \citep{Borucki:2010, Koch:2010, Caldwell:2010}.  Transit durations typically last a few hours, and they are separated by intervals of days to years. The Kepler mission collects data mostly on a 30-minute cadence near-continuously with $>$ 92\% completeness. Detection of transits by small planets requires parts-per-million photometric precision \citep{Jenkins:2002}.  The primary objective will be realized through the combination of space-based photometric sensitivity, regular observing cadence, a 116 square degree field-of-view containing a large number of target stars, and high duty cycle \citep{Koch:2010}.

One of the primary Kepler legacies is a community archive containing the multi-year, time-series photometry of $1.5 \times 10^5$ astrophysical targets observed for the planetary survey \citep{Batalha:2010} and community-nominated targets of other astrophysical interest. In addition to exoplanet science,  Kepler provides unique datasets and raises scientific potential in fields such as asteroseismology \citep[e.g.][]{Bedding:2011,Chaplin:2011,Antoci:2011,Beck:2011}, gyrochronology \citep{Meibom:2011} , stellar activity \citep[e.g.][]{Basri:2011,Walkowicz:2011}, binary stars \citep[e.g.][]{Carter:2011,Derekas:2011,Slawson:2011,Thompson:2012}, and active galactic nuclei \citep{Mushotzky:2011}.  Our aim for this paper is that it is used as a reference guide for the astronomical 
community who wishes to use Kepler data.  This paper will address how the community can mitigate and exploit the public data for stellar and extragalactic science.  Many of the techniques presented here are applied to archived light curves and target pixel files and will help optimize the data for astrophysical research.  

An understanding of the nature of archived data is critical for the effective exploitation of the Kepler legacy.  Kepler provides high-precision photometry on the 1 and 30 minute timescales.  The data also contain artifacts that occur through spacecraft operation events and systematic trends over longer timescales as a natural consequence of mission design \citep{vanCleve:2009,Christiansen:2011a}. Artifacts can both mask astrophysical signal and be misinterpreted as astrophysical in origin \citep{Christiansen:2011a}. Furthermore, cavalier approaches to artifact mitigation (e.g. using simple function fits to time-series data) can destroy astrophysical signal. Archived Kepler data have been processed through a data reduction pipeline \citep{Jenkins:2010a}. The pipeline functions include pixel-level calibration \citep{Quintana:2010}, simple aperture photometry \citep{Twicken:2010a} and artifact mitigation \citep{Twicken:2010b}. This third step of artifact removal will always be a subjective process. An archive user can either work with the default pipeline correction to be of suitable quality to enable their scientific objectives or choose to perform artifact removal themselves, starting from either the calibrated pixels or aperture photometry.

In this paper we provide a guide to Kepler data, insight into the nature of systematic artifacts, a description of how to remove them manually to best effect, and working examples using open source data analysis software. In the next section, we list the available data and documentation. We describe pixel level data in Section 3 and how they can be 
employed to construct new light curves and mitigate systematic artifacts. Section 4 focuses on the exploitation and 
limitations of Kepler's aperture photometry products. In Section 5, we introduce cotrending basis vectors, which can be used to remove instrumental artifacts from time-series aperture photometry. The related problem of stitching multiple quarters of data together is discussed in Section 6. The appendix of this paper contains three worked examples of how to mitigate specific common issues with archived Kepler data. We conclude the paper with a listing of helpful resources\footnote{Kepler archive users can find documentation, data analysis software and helpdesk support at http://keplergo.arc.nasa.gov.} for pursuing further artifact mitigation. 

\section{Kepler Summary and Resources}
In order to maintain long-term stable pointing upon a single field, the Kepler spacecraft is in an Earth-trailing, 372 day heliocentric orbit. Kepler's 0.95-m aperture Schmidt telescope carries a photometer with an array of 42 CCD chips \citep{Koch:2010}.  Each CCD has a direct neighbor, and collectively they are referred to as a ``module'', of which there are 21. Each module has 4 output nodes. Therefore the CCD array and targets falling upon silicon are often mapped according to module and output numbers. Alternatively, mission documentation also refers to output nodes by ``channel'' number, which ranges from 1--84. The module, output and channel locations are provided for each target within the archived meta-tables at the Mikulski Archive for Space Telescopes (MAST)\footnote{http://archive.stsci.edu/kepler}. The Kepler Instrument Handbook \citep{vanCleve:2009}  maps module, output and channel numbers to the detector array.  The field of view and pixel scale were designed to maximize the number of resolvable stars brighter than $K_p$ = 15. $K_p$ refers to an AB magnitude \citep{Oke:1974,Brown:2011} across the Kepler 425--900 nm bandpass \citep{vanCleve:2009}.  The $K_{P}$ magnitude is composed of the star's calibrated {\it g}, {\it r}, and {\it i} magnitudes, obtained during the pre-launch ground based survey that constructed the Kepler Input Catalog \citep{Brown:2011}. 
The Kepler field of view spans 115.6 square degrees over 94.6 million $3.98 \times 3.98$ arcsec detector pixels, with 95\% of the encircled energy is contained within 3.14-7.54 pixels \citep{vanCleve:2009}.   
The Kepler field contains 10 million stars brighter than the magnitude limit of $K_p$ = 20--21. The camera takes one 6.02s image across the full field every 6.54 s. Exposures are summed onboard and stored at either long cadence (29.4 min; see \citet{Jenkins:2010b}) or short cadence (58.85 s; see \citet{Gilliland:2010}). Science data are downloaded approximately once per month when Kepler leaves the field temporarily to point its high-gain antenna towards Earth \citep{Haas:2010}. The number of pixels collected and transmitted is a trade-off between maximizing the number of targets delivered and minimizing the length of the data gaps. For long cadence observations, a maximum of $5.4 \times 10^{6}$ pixels are stored onboard in the spacecraft Solid State Recorder \citep{Jenkins:2010a}, and the number of targets can range from 150,000 to 170,000. Short cadence data are limited to 512 targets. Download requires approximately a 24-hour hiatus in data collection. 

The long and short cadence pixels equate to less than 6\% of the detector plane, and the remaining pixel data are not stored. The stored pixels are chosen strategically to provide postage stamp images centered on the positions of Kepler targets \citep{Batalha:2010}. The size of a postage stamp increases with target brightness, and the yield lies typically between 163,000--170,000 targets per month. The critical concept for understanding instrumental artifacts is that in order to maximize the number of targets collected, the postage stamp sizes and shapes are chosen to maximize the per-target photometric signal-to-noise on the 3--12 hour timescales of exoplanet transits. The postage stamps do not contain all the flux from a target because the collection of the target's faint PSF wings degrades signal-to-noise by the inclusion of more sky background. The pixels within a postage stamp are defined by a calculation that combines the photometry and astrometry within the Kepler Input Catalog \citep[KIC;][]{Brown:2011} and an analytical pixel response model for the detector and optics \citep{Bryson:2010b}.  The postage stamps are fixed within the pixel array; they do not evolve over a 93-day observation period, or ``quarter''.  A new target list is uploaded to the spacecraft after each quarterly roll.  Changes in the target list occur due to new detector geometry, operational developments to the exoplanet survey, and community-led science programs.  Any time-dependent variation in the position of the target or the size of the point spread function will result in a redistribution of flux within the postage stamp pixels. The spacecraft pointing stability is good to typically 20 milliarcsec over 6.5 hours but the high precision light curves can contain systematic noise that manifests from thermally-driven focus variations, pointing offsets, and differential velocity aberration \citep{Christiansen:2011a, vanCleve:2009}. Many of the systematics within the archived light curves are the result of time-dependent light losses as the target wings fall out of the pixel apertures and time-dependent contamination by neighboring sources falling into the pixel apertures.  All collected data are stored and propagated to the community by the MAST. Technical manuals and reference material describing the mission, its data products and defining mission-specific terminology are also archived at the MAST.

\begin{itemize}
\item {\bf Kepler Instrument Handbook} - describes the design, operation, and in-flight performance of the Kepler spacecraft,  telescope, and detector \citep{vanCleve:2009}.

\item {\bf Kepler Data Processing Handbook} - provides a description of the algorithms and pipelined data processing performed upon collected data \citep{Fanelli:2011}.

\item {\bf Kepler Archive Manual} - describes the content and format of archived Kepler data products and the available archive search and retrieval resources \citep{Fraquelli:2011}.

\item {\bf Kepler Data Characteristics Handbook} - defines the causes and provides illustrative examples of characteristics within the Kepler time-series data and systematic artifacts found therein \citep{Christiansen:2011a}.

\item {\bf Kepler Data Release Notes} - provide an impact assessment of systematic artifacts and spacecraft events upon Kepler time-series data.

\end{itemize}

Within the MAST archive, Kepler data are stored as files designed to the format and conventions of the Flexible Image Transport System \citep[FITS;][]{Pence:2010}. Short cadence target data files each contain one month of observations, and long cadence data files contain one quarter of observations. Instructions for the MAST data search and retrieval tools are provided in the Kepler Archive Manual. The three primary forms of Kepler science data stored within the archive are:

\begin{itemize}

\item {\bf Full-frame images (FFIs)} - using all the pixels on each detector channel, the Kepler spacecraft accumulates one 29.4 min image comprised of a sequence of 270 consecutive 6.02s exposures coadded onboard, once per month, before each data download. The 84 channels are stored within a single FITS file. FFIs are collected primarily for engineering purposes but provide a scientific resource in their own right -- high-precision photometry of the entire Kepler field of view on a 1-month cadence. Additionally, FFIs can be employed to assess the target's flux level and sources of contamination from nearby objects within a target's pixel aperture. 

\item {\bf Target Pixel files (TPFs)} - each TPF stores a time-stamped sequence of uncalibrated and calibrated postage stamp pixel images of a Kepler target over a long cadence quarter or short cadence month. The TPF content is discussed further in the context of systematic artifacts in Section 3.

\item {\bf Light curve files} - each file contains time-series photometry for an individual Kepler target, derived from an optimized subset of pixels contained within the associated TPF. A more detailed examination of these files and the consequences of the systematic effects within the TPFs are provided in Section 4.

\end{itemize}

\section{Kepler Archived Target Pixel Files (TPF)} 

For each individual Kepler target, light curve files (discussed in Section 4) are accompanied by a TPF. The TPF is the single-most informative resource in the archive for understanding the instrumental, non-astrophysical features within a target light curve. Consequently, it is recommended that users always examine a TPF in conjunction with its light curve. They provide the detector pixel and celestial coordinate mapping of a target mask and its subset containing the optimal aperture. TPFs reveal the motion of a target across the optimal aperture, and the motion of nearby contaminating sources.  

As with the light curve file, the TPF content is organized by timestamp. The timestamps are the barycentric Julian date at the midpoint of each accumulated exposure. Pixel data are presented as a time-series of photometrically calibrated images, one image per timestamp, all located within the first extension of the FITS file. Typical Kepler targets ($K_p > 12$) will contain 10--50 pixels, but bright sources will include a larger pixel set. A typical quarter will contain approximately 4,300 collected images in long cadence for each target, while a typical month will yield 43,200 images in short cadence mode. For each timestamp within the TPF, there is an image of the uncalibrated pixels collected around a target. This group of pixels is referred to as the {\it target mask}, which contain pixels assigned either to the optimal aperture or a halo around it. The {\it optimal aperture} is the set of pixels over which the collected flux is summed by the Kepler pipeline to produce a light curve. The {\it halo pixels} are those pixels surrounding the optimal aperture pixels, which are used for calibration purposes and provide operational margin. 

Additionally, for each timestamp, the TPF includes the raw counts (FITS column labeled ``RAW\_CNTS'') a fully calibrated postage stamp pixel image (FLUX) which incorporates bias correction, dark subtraction, flat fielding, cosmic ray removal, gain and nonlinearity corrections, smear correction (the Kepler photometer has no shutter), and local detector electronic undershoot (i.e. sensitivity of the pixel response to bright objects). The Data Processing Handbook \citep{Fanelli:2011} and \citet{Caldwell:2010} contain further details of these corrections.  The TPFs also provide images for each timestamp containing the 1-$\sigma$ flux uncertainties of each pixel (FLUX\_ERR), a calibrated sky background (FLUX\_BKG), 1-$\sigma$ uncertainties to the sky background (FLUX\_BKG\_ERR), and cosmic rays incidences (COSMIC\_RAYS).  All of these postage stamp images are found in the first extension of the FITS file.  Sky background generally cannot be well-estimated from the TPFs themselves because target masks do not encompass enough sky for a good background estimation. Instead 4,464 pixels across each channel are recorded at long cadence specifically to measure the sky background. Those measurements are interpolated across each target mask to characterize the local sky background for each aperture. 

Each timestamp has a quality flag coupled to it which alerts the user to phenomena and systematic behavior that may bring the quality of the photometric measurement within that timestamp into question. Detrimental behavior is generally associated with one of the following: i) physical events such as cosmic rays followed by short term detector sensitivity dropouts, ii) foreign particles such as dust, iii) spacecraft motion due to attitude-control reaction wheel resets, zero-torque crossing events, spacecraft pointing offsets and/or loss of fine-pointing upon guide stars, iv) time-dependent variation in incident solar radiation and telescope focus (caused by either Kepler's orbit, autonomous operational commands directing the spacecraft to re-point towards a safe direction, or pointings towards Earth for the transmission of data), and v) differential velocity aberration (DVA) which is caused by the spacecraft's orbital motion constantly changing the local pixel scale and field distortion. See \citet{Christiansen:2011a} for more detailed descriptions of each event listed above and \citet{Fraquelli:2011} for a list of quality flags in the TPF and its corresponding events. Note that the list provided here contains only the major causes of artifacts common to many operational quarters of data. Other events do exist and are described in \citet{Christiansen:2011a}.

A pixel bitmap indicating the use of each pixel within the target mask in the Kepler pipeline is stored as an image in the second FITS extension of the TPF. The target masks do not track and follow stellar motion, and the large pixel scale undersamples the point spread function. Optimal apertures in general cannot provide absolute photometry because target flux is always lost outside the pixel borders and contamination from nearby objects falls inside the collection area. Spacecraft motion and focus variation are detrimental to Kepler photometry because time-dependent variation in either property results in different fractions of target light being lost from the aperture and different amounts of source contamination falling into the aperture. Despite Kepler's cadence-to-cadence pointing generally being stable at the milli-arcsecond level, motion larger than this threshold has a measurable consequence at the few $10^{-5}$ photometric accuracy.  Kepler simple aperture photometry is consequently a combination of astrophysical signal from the target and systematics from the spacecraft and its environment. The more precise the scientific requirement, the more care one must take to achieve accurate photometric results. 

Using the motion across the detector of a set of reference stars, the Kepler pipeline predicts the motion of the target over time and provides a predicted position for each timestamp within the TPF (the FITS columns labelled POS\_CORR1 and POS\_CORR2).  Predictions of the target motion trace many of the Kepler systematics, and this measured astrometric deviation from the reference star predictions are likely to be astrophysical in nature.  For example, as a target's brightness varies, the centroid of the flux distribution across the pixels will move if there are contaminating sources in the target mask's pixels.  The change in the flux centroid position is a method that can help detect faint, unresolved background binaries or other variable stars, which can appear as a false detection of planetary transit. (e.g. see Appendix A).

\begin{figure}
\includegraphics[scale=0.6,angle=90]{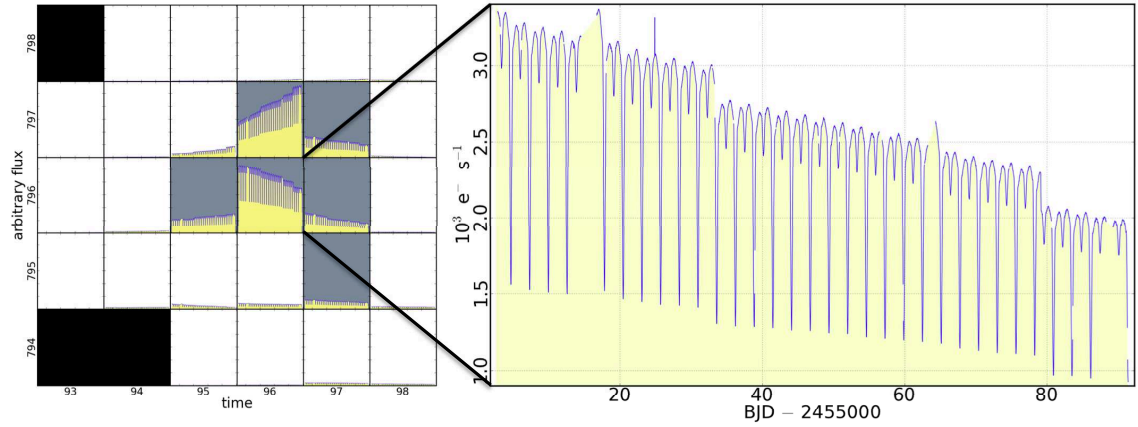}
\caption{Example output plots from PyKE tools {\it keppixseries} and {\it kepmask}. (left) The plot is created from the target pixel files. The numbers along the axes identify the pixel column and row on the CCD.  The pixels comprising the optimal aperture for the target have a gray background.  No data are collected in the black pixels, and the white pixels, which do sometimes typically collect some of the target's flux, are the halo pixels (see Section 3 for definitions).  (right)  The axes labels ``arbitrary flux'' and ``time'' refer to the photometric time series plotted for each individual pixel.  Each light curve in the gray and white pixels is plotted on an identical linear flux scale.  The target flux is continuously being redistributed among the neighboring pixels of the optimal aperture as the quarter progresses.  The plot on the right also shows the decline of target flux, which indicates the target moved within the optimal aperture pixels.  The missing flux was redistributed to the pixel above, as seen in the left panel.  The pixel light curves are for Quarter 2 long cadence observations of V1950 Cyg.}
\label{q2-tpf}
\end{figure}

Figure \ref{q2-sap} provides a typical example of a Kepler long cadence light curve, specifically the quarter 2 simple aperture 
photometry of the eclipsing binary star V1950 Cyg.  In this example, the flux affected by short-term systematics are dominated by the high-amplitude intrinsic variability of the target, but the effects of DVA still clearly manifest as a 3\% decrease in target flux over the duration of the quarter.  The situation is clarified by inspection of the associated TPF.  Figure \ref{q2-tpf} reconstructs the photometry for each individual pixel within the target mask of the source of interest.  The mask includes both the star itself and the halo pixels around the star.  The pixels in Figure \ref{q2-tpf} with gray backgrounds are chosen by the Kepler pipeline to be the photometric optimal aperture for the archived light curve.  The other pixels in the halo are associated with the target mask but they play no part in the calculation of the archived time-series photometry.  The wings of the point spread function extend into many pixels surrounding the optimal aperture.  Kepler apertures, in general, are designed to maximize signal-to-noise which usually undercaptures target flux.  Without full capture of a source, motion and focus-induced artifacts in the summed light curve are inevitable.   In Figure \ref{q2-tpf}, as the quarter proceeds, note how the flux captured in each pixel can increase or decrease as the target moves across the pixel array.  This example shows that the flux is continuously being redistributed between neighboring pixels of the optimal aperture, and different amounts of flux will fall outside this aperture over time. The times and nature of events related to the discontinuities seen in the TPF light curves are recorded under the QUALITY column in the TPF FITS headers.

The example of V1950 Cyg illustrates how Kepler light curves can easily be affected by instrument systematics (e.g., DVA).  There are instances where extracting target pixels over a larger detector area compared to the optimal aperture chosen by the pipeline minimizes or removes the instrument effects from the light curves.  However, the artifact mitigation of including more pixels comes at a cost.  More pixels within the flux summation decreases the photometric signal-to-noise by adding more sky background from the additional pixels.  The user must decide whether adding more pixels is an acceptable mitigation for the systematics.  Many Kepler targets exist within crowded fields, and multiple nearby sources may contribute to the flux within the target mask.  While it is sometimes beneficial to include more pixels within a modified optimal aperture, in the worst case scenario, the systematic errors in light curves and source contamination can increase significantly by including more pixels in extraction.

Inspection of the time-series images within the TPF can help reveal whether photometric variability is due to source contamination or is intrinsic to the target.  An example of a background object contaminating the light curve of a target is provided in Appendix A.

\section{Kepler Light Curve Files} 

The Kepler archive stores target specific light curve files in binary FITS format that have been derived from each TPF. The format of the FITS file is defined in \citet{Fraquelli:2011}. The time stamps, quality flags, predicted target motion relative to the detector pixels, and pixel bitmap information described in Section 3 are copied into the light curve FITS tables. Kepler light curve files contain a number of columns containing flux information.
Two columns contain simple aperture photometry (SAP) flux with 1-$\sigma$ statistical uncertainties, while a more processed version of SAP with artifact mitigation included called Pre-search Data Conditioning (PDCSAP) flux \citep{Smith:2012} and its uncertainties, populate two more columns.  The sky background values, summed across the optimal aperture, and its 1-$\sigma$ uncertainty are calculated from the TPF and added to the light curve files.  The last set of columns in the light curve FITS files are the timestamped moment-derived centroid positions of the target, as calculated from the calibrated TPF images.   The Data Processing Handbook \citep{Fanelli:2011} provides details on how the centroid positions were calculated.  The centroids are provided in detector pixel row and column coordinates.  The centroid positions can be used as a direct comparison to the motion predicted from a set of reference stars per CCD channel. The purpose of the comparison between measured flux centroid and the centroid predicted from the motion of reference stars is to identify times when these quantities are uncorrelated. Potentially, uncorrelated centroid structure identifies events in the light curve that are caused by fractional changes in contamination from sources close to or unresolved from the target. 

\subsection{Simple aperture photometry (SAP)}

The SAP light curve is a pixel summation time-series of all calibrated flux falling within the optimal aperture, as stored and defined in the TPF. The 1-$\sigma$ errors are calculated from standard Gaussian error propagation of the TPF errors through the sum. Data archive users need to be aware that a SAP light curve can be contaminated by astrophysics from neighboring sources. One can inspect the concurrent TPF to identify contamination.  A new SAP light curve can be extracted from the TPF using a custom selection of pixels, as shown in Appendix B.

Archive users must expect, {\it a posteriori}, that SAP photometry is contaminated by the systematics discussed in Section 3. To continue using SAP data for scientific exploitation, one must decide whether the artifacts will impact their results and conclusions. There is ``low-hanging fruit'' that has dominated Kepler astrophysics activity in the early phases of the mission because the SAP data has proved to be adequate for specific science goals without artifact mitigation. For example, asteroseismology of solar-like oscillations, $\delta$ Scuti and $\gamma$ Doradus pulsations have been hugely successful because signals of frequency $\gtrsim$ 1-d$^{-1}$ are mostly unaffected by the majority of artifacts \citep{Balona:2011a, Uytterhoeven:2011,Balona:2011b}. Some high frequency artifacts that could prove problematic to these programs can be filtered out of the time-series using the quality flags provided. Data analysis of cataclysmic variables, RR Lyr stars and Cepheids are just as successful. While many of the astrophysical frequencies of interest in these pulsators can be longer than a few days and similar to the thermal resettling times of the spacecraft after a pointing maneuver, the large amplitude of target variability dominate over systematics that can consequently be neglected (e.g. \citet{Still:2010, Nemec:2011, Szabo:2011}). 

There are many astrophysical applications that are less likely to benefit from direct employment of SAP data. These include any science relying on more subtle light curve structures and periods longer than a few days, in which case the systematics discussed are more likely to be significant. Investigations of magnetic activity, gyrochronology, binary stars and long period variables must scrutinize the SAP data with great care before proceeding and will most likely benefit from one of three available artifact mitigation methods. The three methods are to use archived PDCSAP photometry (Section 4.2), to re-extract the SAP light curve over a larger set of pixels (Appendix B), or to 
perform a custom correction on the archived SAP data using cotrending basis vectors (Appendix C). These methods and their precise 
application are very subjective. The highest quality Kepler research will in most cases result from the experience and understanding gained by applying all three of these methods to the archived data.

\subsection{Pre-search data conditioning simple aperture photometry (PDCSAP)}

The PDCSAP data included within the archived light curve files are produced by a pipeline module that remains under continuing development at the time of writing \citep{Smith:2012,Stumpe:2012}.  Versions of PDCSAP artifact mitigation algorithms (before November 2011) employed the removal of analytical functions while correlating the photometry with spacecraft diagnostic information such as the focal plane temperature. This approach focused upon solving the problem for the effective detection of exoplanet transits, without discrimination removing both systematic and astrophysical variability that would interfere with transit detection. PDCSAP data provided before November 2011 should not be used without skepticism for any purpose other than transit detection. It is recommended that this older version of PDCSAP data not be used for stellar or extragalactic astrophysics. All such data have been reprocessed by the pipeline with different artifact mitigation algorithms which are more robust for astrophysics and re-delivered before May 30, 2012.

\begin{figure}
\includegraphics[width=\textwidth]{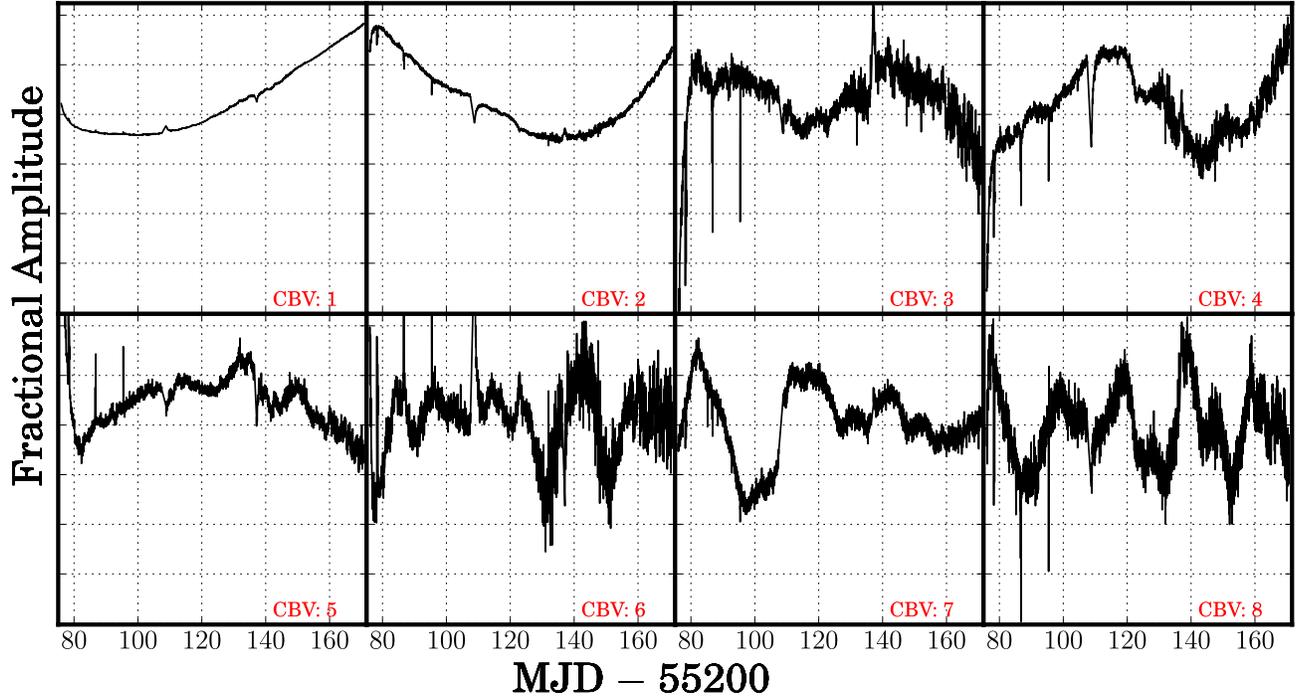}
\caption{An example of eight cotrending basis vectors with the highest principle values, or contribution to systematic variability, from channel 50 over operational quarter 5. Each basis vector is on the same relative flux scale, centered about 0.0 e$^-$ s$^{-1}$.  Basis vectors can be linearly-fit to a light curve and subtracted off to mitigate for systematic effects.  The fit coefficients can either be positive or negative. They run from left-to-right, top-to-bottom, in order of significance.}
\label{cbvs}
\end{figure}

Post October 2011, quarterly data deliveries to the archive were constructed using versions of the pre-search data conditioning (PDC) pipeline module developed to better remove systematic artifacts while conserving more astrophysical signal. This development provides a change of approach and comes in two parts. In the first part, systematic artifacts are characterized by quantifying the features most common to hundreds of strategically-selected quiet targets on each detector channel. For each channel and each operational quarter, this characterization is stored as 16 best-fit vectors called ``Cotrending Basis Vectors'' (CBVs). The basis vectors archived represent the most common trends found over each channel. The CBVs are ranked by order of the relative amplitude they contribute to systematic trends across a channel. An example of the 8 most dominant CBVs for CCD channel 50 over quarter 5 is provided in Figure \ref{cbvs}.  A description of the pipeline algorithm for constructing CBVs is described in \citet{Smith:2012} and \citet{Stumpe:2012}.

In second part of the PDC pipeline module, systematics are removed from SAP time-series by subtracting the CBVs. The results are stored in the archived files and labeled PDCSAP data. The correction is unique to each target.  A weighted normalization for 
each basis vector in the calculation is determined by fitting basis vectors to the SAP data, but the CBV weighting and ``best'' astrophysical solution remains a subjective problem. The process is therefore repeatable by archive users and tunable. The pipeline has configured the tuning to provide the most effective conservation of astrophysics within the Kepler targets each quarter by detector channel as a statistical sample. The pipeline algorithm therefore provides a significant improvement in the quality of artifact mitigated photometry. However the PDC algorithm is not tuned to individual targets or specific classes of target. As the PDC pipeline continues to mature, the number of individual problematic cases in the archive will shrink. However, for any individual target, we recommend direct comparison of the three artifact mitigation methods available in order to understand whether the archived data provides a solution optimized to the users' scientific requirement. A manual re-extraction of a target light curve from a TPF will produce a SAP time-series, but not a PDCSAP time-series. If artifact mitigation is required subsequent to light curve extraction, then the only viable option is to manually fit the CBVs. Manual CBV fitting and subtraction is the subject of Section 5. 

\section{Removing systematic artifacts with cotrending basis vectors} 

\begin{figure}
\includegraphics[width=\textwidth]{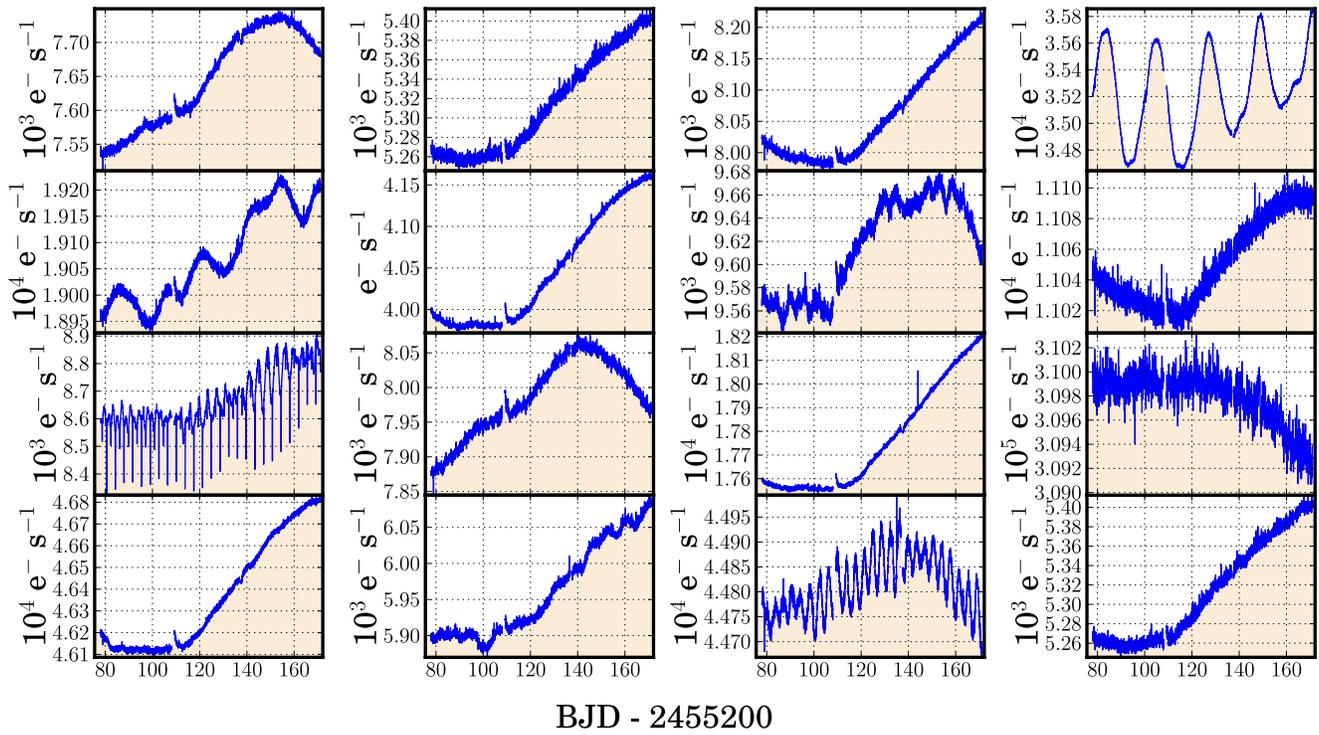}
\caption{Sixteen long cadence light curves chosen at random from quarter 5, channel 50. All light curves show some degree of correlation. The most obvious common feature is an increase in the flux level over the course of the quarter. This is the manifestation of differential velocity aberration.}
\label{dva}
\end{figure}

\begin{figure}
\includegraphics[width=\textwidth]{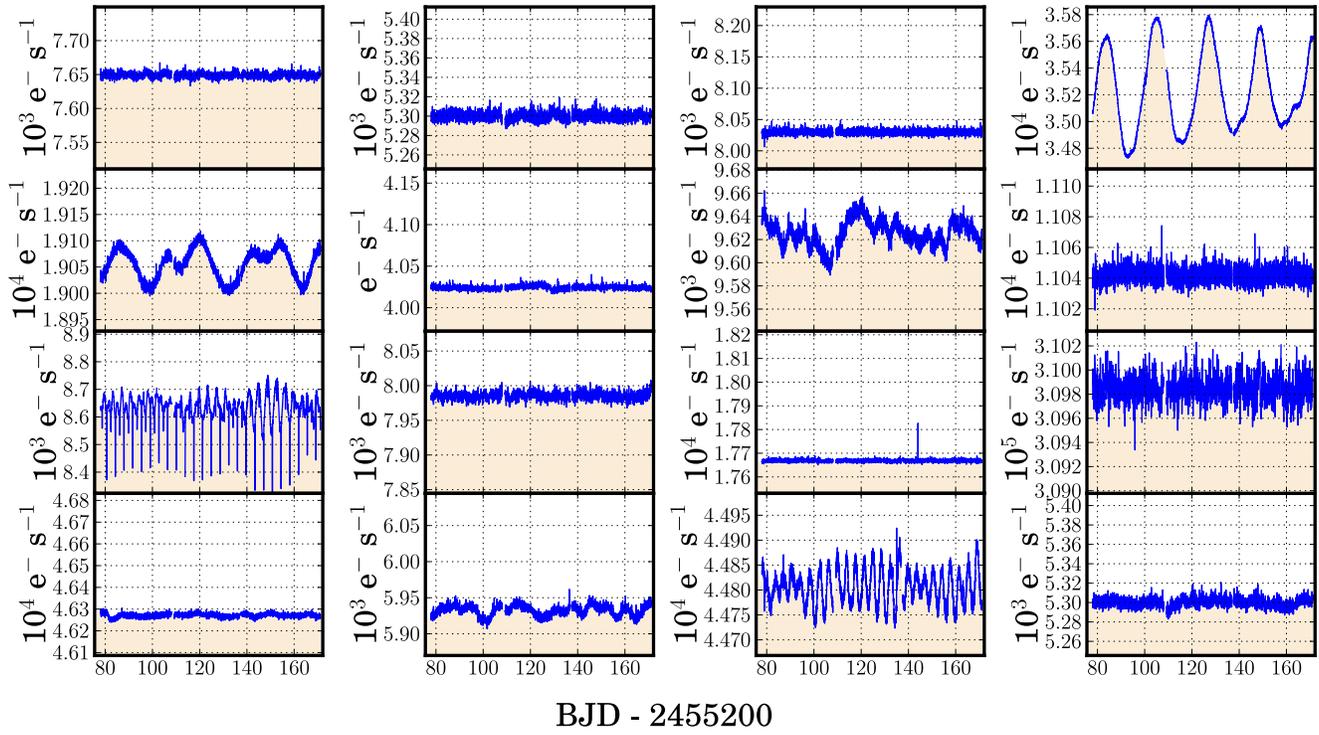}
\caption{The sixteen quarter 5 SAP light curves presented in Figure~\ref{dva} after the best-fit CBV ensemble has been subtracted.}
\label{dva-corr}
\end{figure}

The 16 most significant CBVs per channel for each quarter are calculated by the Kepler pipeline and packaged as FITS binary files. These CBVs are available for download at the MAST\footnote{http://archive.stsci.edu/kepler/cbv.html}. File content and format are described in \citet{Fraquelli:2011}. With the provision of CBVs, the responsibility rests with the archive user to either improve upon the existing artifact mitigation done by the pipeline or perform manual artifact removal from photometry re-extracted from a TPF. Basis vectors are usually fit to the SAP light curve linearly, i.e. each basis vector is scaled by a coefficient and subtracted from the flux time-series. Computationally, the most efficient method is a linear least-squares fit. In Figure \ref{dva} we plot 16 SAP light curves from channel 50 in quarter 5. In Figure \ref{dva-corr}, we plot the same light curves of Figure \ref{dva} but with the most significant CBVs fit and subtracted. We can see that in all the light curves in Figure \ref{dva-corr}, the systematic trends have been greatly reduced and the astrophysics is more clearly delivered.  Appendix C provides an example of how to apply the CBVs to data.

An important decision for the CBV user is how many basis vectors to fit and remove from the SAP data. Fitting too few will capture instrumental artifacts less effectively. However, using too many can overfit the data, removing real astrophysical features. A further consideration is that no basis vector is perfect. The inclusion of each additional CBV to the fit adds a noise component to the data. The choice of CBV number is in reality a trade between maximizing the removal of systematics on the one hand, and avoiding the removal of real astrophysics and minimizing the effects of CBV noise on the other. A minimum of two basis vectors should be fit to the data because, instead of strictly enforcing a constant first or second basis vector, CBVs are created by mixing a constant offset with the strongest non-constant basis vector. We find that an interative method is the most effective approach, starting with two basis vectors and increasing the number of vectors monotonically until deciding upon a subjective optimal fit.  Appendix C shows example fits of 2, 5, and 8 CBVs to demonstrate the iterative method of determining the number of CBVs to use.

Occasionally the linear least-squares fit is not sufficient, and a more robust fitting method must be utilized. One option is to fit the CBVs to the SAP time-series using an iterative-clipping least-squares method. This method identifies data points outside of a distance threshold from the best fit. Data points outside of the threshold are excluded and the fit recast. This procedure iterates until no further data points are rejected and is more robust to outliers than a regular least-squares fit. Alternatively, rapid, high amplitude astrophysical variability can bias the goodness of fit away from the best astrophysical solution. 

Some sources of astrophysical variability, such as large amplitude, semi-regular variable stars, cannot be corrected satisfactorily with CBVs. Such sources which vary on a similar timescale to the length of a quarter are particularly problematic because the astrophysical signal has the same frequency as the most dominant basis vector. In addition, if the astrophysics is too similar in structure to the trends created by differential velocity aberration, cotrending corrections with CBVs may not be adequate. In these cases, we do not recommend using the CBVs to mitigate for long-term artifacts. 

It should be noted that the CBV method for removing systematic trends relies on there being a large number of stars on a channel to well-describe the systematic effects present in the data. There are only 512 stars observed in short cadence mode at a given time, thus there are not enough stars on a single channel to fully capture the systematics present on 1-min cadence. The method currently available for mitigating short cadence artifacts is to interpolate the long cadence basis vectors over the short cadence timestamps. Artifacts on timescales less than 30-min cannot be removed from short cadence data using the CBV method.

\section{Stitching Kepler quarters together} 

\begin{figure}
\includegraphics[scale=0.6,angle=0]{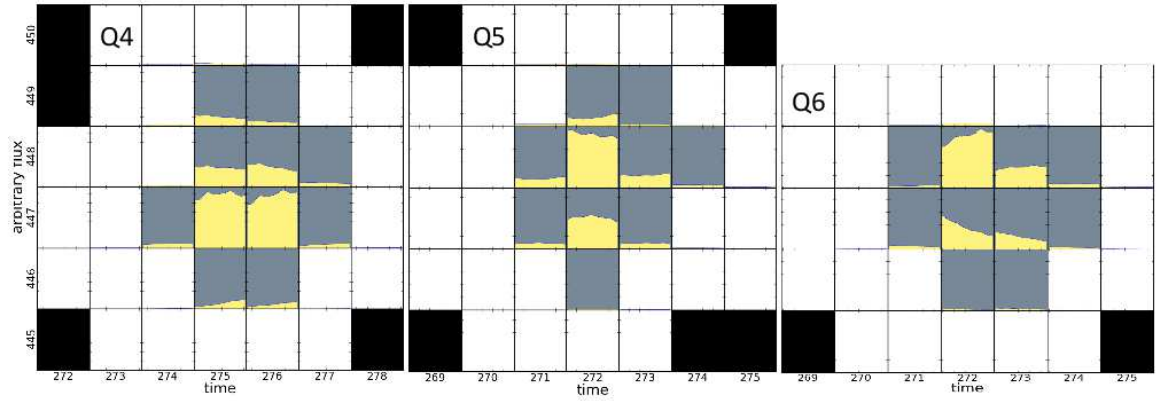}
\caption{Light curves extracted from all single pixels within the quarter 4, 5 and 6 Target Pixel Files for KIC 9603833 (the symbiotic star StHA 169). Gray pixels comprise the optimal apertures that yield the archived light curves. The target point-spread function is distributed across neighboring pixels differently from quarter to quarter and hence the optimal aperture varies in size and shape from quarter to quarter. The amount of target flux falling outside of the optimal aperture is quarter-dependent. Data are not collected from the black pixels. The plots were created with the {\it keppixseries} PyKE task, using the {\it plottype=global} option.}
\label{stha169-pix}
\end{figure}

In Section 3, we described how systematic photometric artifacts result directly from the motion of targets within their pixel apertures due to differential velocity aberration, spacecraft pointing, and focus variations. Similar in nature, Kepler data will often exhibit discontinuities across the data gaps that coincide with the quarterly rolls of the spacecraft. After each roll maneuver, most Kepler targets fall on a different CCD channel and the target's point-spread function will be distributed differently across neighboring pixels. Naturally this redistribution requires a new computation of the target mask and optimal aperture size, taking into account the point-spread function, new CCD characteristics, and new estimates of nearby source crowding. The operational outcome is often a different mask shape, with differing amounts of flux within the optimal aperture from both the target and contaminating sources. An illustrative example of this problem is provided by the quarter 4-6 calibrated pixels collected for the symbiotic star StHA 169 (KIC 9603833), plotted in Figure \ref{stha169-pix}. 

\subsection{Suggested stitching methods}
There are several methods available to attempt correcting for photometric discontinuities across the quarter gaps. None of the methods are well-suited to all occasions. Individually they can perform good corrections under specific conditions.  We discuss the three methods below.

\subsubsection{Crowding and aperture flux loss adjustment}
The first method is to align different quarters using the time-invariant approximations for crowding and aperture flux losses stored within the light curve FITS file keywords. These unitless keywords are stored in all Kepler data processed after September 2011. For earlier data the same quantities are populated in the meta-tables of the data search and retrieval page at MAST. The FITS keyword FLFRCSAP contains the fraction of target flux falling within the optimal aperture. The keyword CROWDSAP contains the ratio of target flux relative to flux from all sources within the photometric aperture. Both quantities are the average value over each quarter and are estimated using point-spread function and spacecraft jitter models \citep{Bryson:2010} combined with source characteristics found within the KIC \citep{Brown:2011}. The PDC time-series data archived within the FITS light curve files have both of these corrections applied by default. Both corrections can be applied to SAP data manually using the {\it keparith} task within the PyKE package (see Appendices for a description of PyKE). 

The limitations of this first method are two-fold. The corrective factors supplied are model-dependent. The characterization of the Kepler point-spread function does not provide the same order of photometric precision as aperture photometry.  Furthermore, PSF modeling of the pixels within an optimal aperture is only as complete as the KIC, which is complete only at $K_{P} < 17$.  Secondly, the corrective factors are averaged over time, whereas in reality they vary from timestamp to timestamp as the field moves within the aperture. Furthermore, as the target and neighboring stars vary independently, the crowding correction in reality is an additive one rather than multiplicative. Therefore, while these two keywords collectively provide the simplest method of quarter stitching, the solution is often inadequate. Given that target images for each timestamp are provided in the archive within the Target Pixel Files, there is some scope for Kepler users to improve upon these corrections by fitting a field and point-spread function model to the individual images within these files. The Pixel Response Function (the combination of point-spread function and spacecraft jitter) information is available at the MAST archive within the focal plane characteristics download table\footnote{http://archive.stsci.edu/kepler/fpc.html}. Fitting a PSF model can yield improvements over the provided correction factors because the time-dependent variations associated with the problem are mitigated, and the user can take individual care characterizing all neighboring sources contributing to flux within the aperture. The primary limitation remaining after these improved steps will be the accuracy of the point-spread function model.

\subsubsection{Normalization of light curves}
A second approach that can yield adequate results is to individually normalize light curves on either side of a quarter gap by a functional fit or statistical measure of the data. Some simple corrections by statistical representations of the data, such as mean, median and standard deviation, are available through the PyKE tools such as {\it keparith}. This method of correction is however a mathematical convenience and users should remain aware of the non-physical biases that they may introduce into the data.

\subsubsection{Using more pixels in the target mask}
\begin{figure}
\includegraphics[scale=0.6,angle=90]{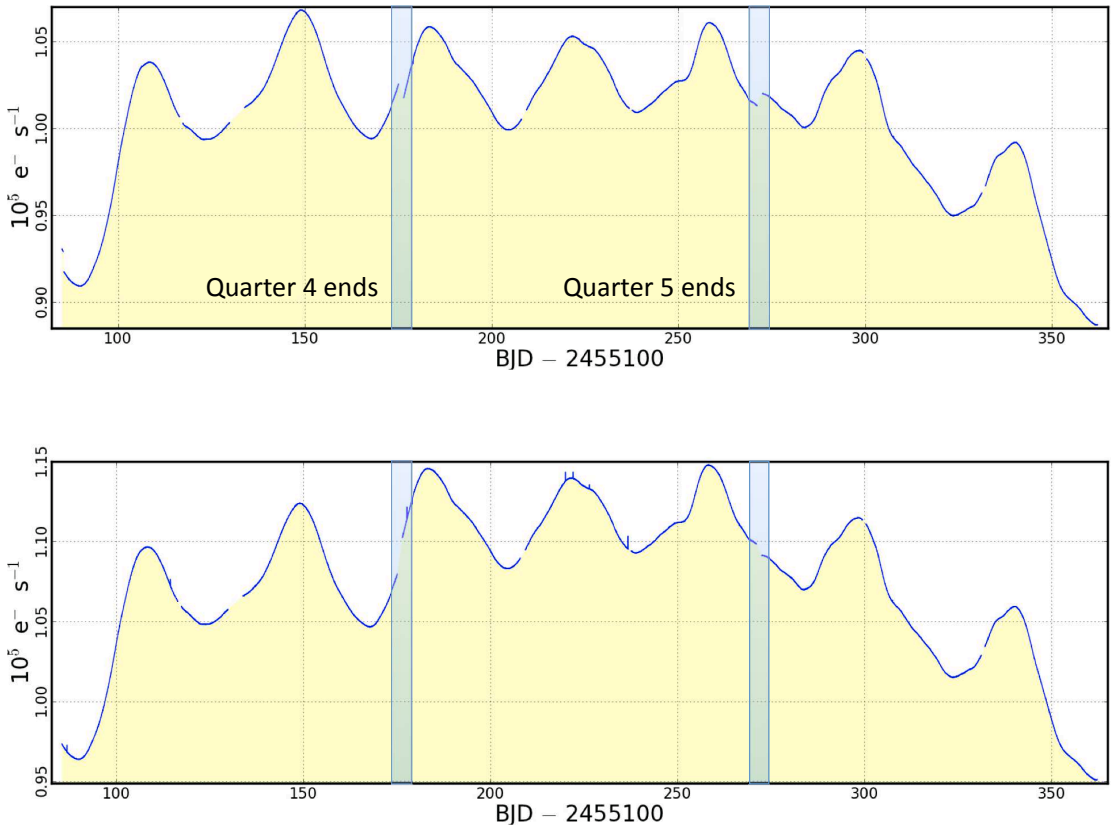}
\caption{Upper: The archived SAP quarter 4, 5 and 6 light curves for StHA 169 (KIC 9603833).  Each light curve was extracted from the optimal apertures (grey pixels) defined in Figure \ref{stha169-pix}. Photometric discontinuities occur at each quarterly roll due to the redistribution of target flux over a new CCD (indicated by the blue bars for clarity), and the redefinition of the optimal aperture. Lower: Quarterly roll discontinuities are reduced by re-extracting the three light curves over all available pixels in the target masks. In this specific example, the redefined optimal aperture collects more of the target's flux without introducing significant contaminating flux from nearby sources. Extraction was performed using the PyKE {\it kepextract} task with the {\it maskfile=all} parameter. Some further examples of pixel extraction are provided in Appendices A and B.}
\label{quartergap}
\end{figure}

As described in Section 3, a third method which often proves to be successful is increasing the number of pixels within the target mask using the PyKE tasks {\it kepmask} and {\it kepextract}. This approach will prove adequate if the optimal aperture can be increased to a large enough size as to make target flux losses out of the aperture negligible, while avoiding significant contamination from nearby sources. This third method will introduce additional shot noise into the resulting light curve by the inclusion of more source and background flux. The example of StHA 169 over quarters 4-6 yields an adequate correction by this method, as demonstrated in Figure \ref{quartergap}

\section{Summary}
This paper has provided a general description of the systematics contaminating archived Kepler data and an introduction to the mitigation of those artifacts. Three detailed examples of artifact mitigation are supplied in Appendices A, B and C. Successful mitigation is not guaranteed in all cases. Conceptual understanding, methodology, and open source software development are maturing such that the quality of mitigated data archive products continues to increase with time.  Several approaches have been developed to manually improve the fidelity of aperture photometry. In order to minimize the impact of aperture photometry artifacts and obtain the highest quality time-series data, archive users are recommended to explore all of the methods discussed in this paper. The most correct approach for individual targets will be subjective and will be achieved through experimentation and hands-on experience.

To date, exploitation of Kepler data has naturally been dominated by its primary mission goals of exoplanet transit detection \citep{Borucki:2010} and asteroseismology of solar-like stars \citep{Chaplin:2011}. By the nature of the mission design, both research areas continue reaping a rich harvest from the Kepler archive.  However, there is a sensitivity threshold for both of these disciplines that will require more state-of-the art artifact mitigation before reaching their full potential.  Similarly Kepler has the potential through multi-year, highly-regular monitoring to provide a startling legacy archive for active galactic nuclei, stellar activity, gyrochronology,  and, perhaps most-compellingly, stellar cycles, for example.  For much of the detailed astrophysics with Kepler data, and for Kepler to reach its broader scientific potential, the challenges of removing aperture photometry systematics must be met.

\acknowledgements \begin{center} {\bf Acknowledgements} \end{center} Funding for the Kepler mission is provided by
NASA's Science Mission Directorate.  The Kepler Guest Observer Office is funded through NASA co-operative agreement NNX09AR96A.  All of the data presented in this paper were obtained from the Multimission Archive at the Space Telescope Science Institute (MAST). STScI is operated by the Association of Universities for Research in Astronomy, Inc., under NASA contract NAS5-26555. Support for MAST for non-HST data is provided by the NASA Office of Space Science via grant NNX09AF08G and by other grants and contracts.

\appendix

\begin{center} {\bf Appendices: Data analysis examples} \end{center}
 
PyKE\footnote{http://keplergo.arc.nasa.gov/PyKE.shtml} is a suite of python software tools developed to reduce and analyze Kepler light curves, TPFs, and FFIs.  PyKE was developed as an add-on package to PyRAF - a python wrapper for IRAF\footnote{http://iraf.noao.edu} that provides for a new tool development to occur entirely in the python scripting language.  PyKE can also be run as a stand-alone program within a unix-based shell without compiling against PyRAF.

We present three general examples of how to prescribe and mitigate source contamination and systematic artifacts within the Kepler data using the PyKE software. These examples provide guidelines for the reader to follow, but data users are encouraged to experiment with the tunable input parameters for each tool. The procedures outlined in this section will not be optimal for all science, and it will ultimately be up to the user to determine what does optimize their scientific return. Ultimately these tools provide the user with flexibility to tune pixel extraction and artifact mitigation to the scientific potential of individual target data.

\section{Example 1: Pixel Images and Source Contamination}

\begin{figure}
\includegraphics[width=\textwidth]{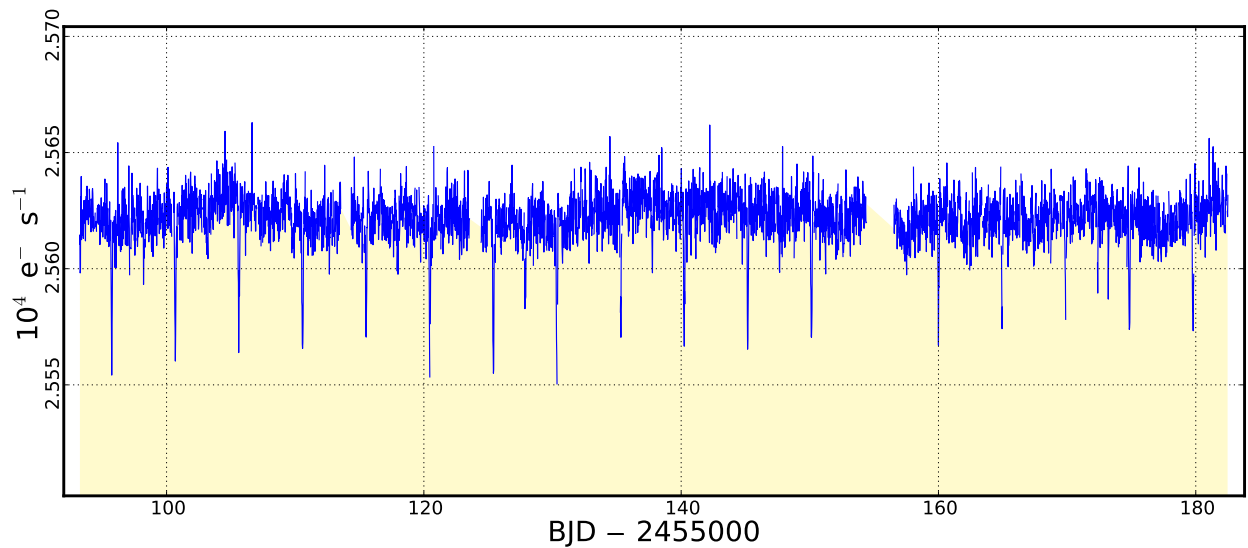}
\caption{The archived quarter 3 PDCSAP light curve of KIC 2449074. The regular dips in brightness every 4.9 d resemble a planetary transit.}
\label{lc-transits}
\end{figure}

Due to the typical angular size of Kepler photometric apertures and the relatively crowded fields within the Kepler field of view, one cannot be certain whether astrophysical variability across a Kepler light curve comes entirely from the target star. The likelihood of source confusion around any given target is high. In order to resolve the sources of variability within a target mask, the archive user should examine the TPF file. For the purposes of this example, the archived Pre-search Data Conditioned light curve of KIC 2449074 is shown in Figure \ref{lc-transits}, as rendered by {\it kepdraw}.  While the PyKE tasks can be operated entirely through GUI-driven operation using the {\it epar} function on the command line, for the sake of clarity in these examples, we provide the command line task invocations within the PyRAF environment:

\begin{quote}
{\tt kepdraw infile=kplr002449074-2009350155506\_llc.fits \\outfile=kplr002449074-2009350155506\_llc.png datacol=PDCSAP\_FLUX ploterr=n errcol=PDCSAP\_FLUX\_ERR quality=y}
\end{quote}

The above command is asking that the PDCSAP\_FLUX column in the archived FITS file kplr002449074-2009350155506\_llc.fits be plotted to a new file called kplr002449074-2009350155506\_llc.png. Plotting of the 1-$\sigma$ error bars from the PDCSAP\_FLUX\_ERR column will be suppressed, and timestamps with non-zero quality flags will be ignored. PyKE will request more parameters through the command line before creating the plot but users can take the default options. Experimentation will reveal that these additional parameters control the look and feel of the plot -- e.g. colors, line widths, fonts, etc. 

\begin{figure}
\includegraphics[width=\textwidth]{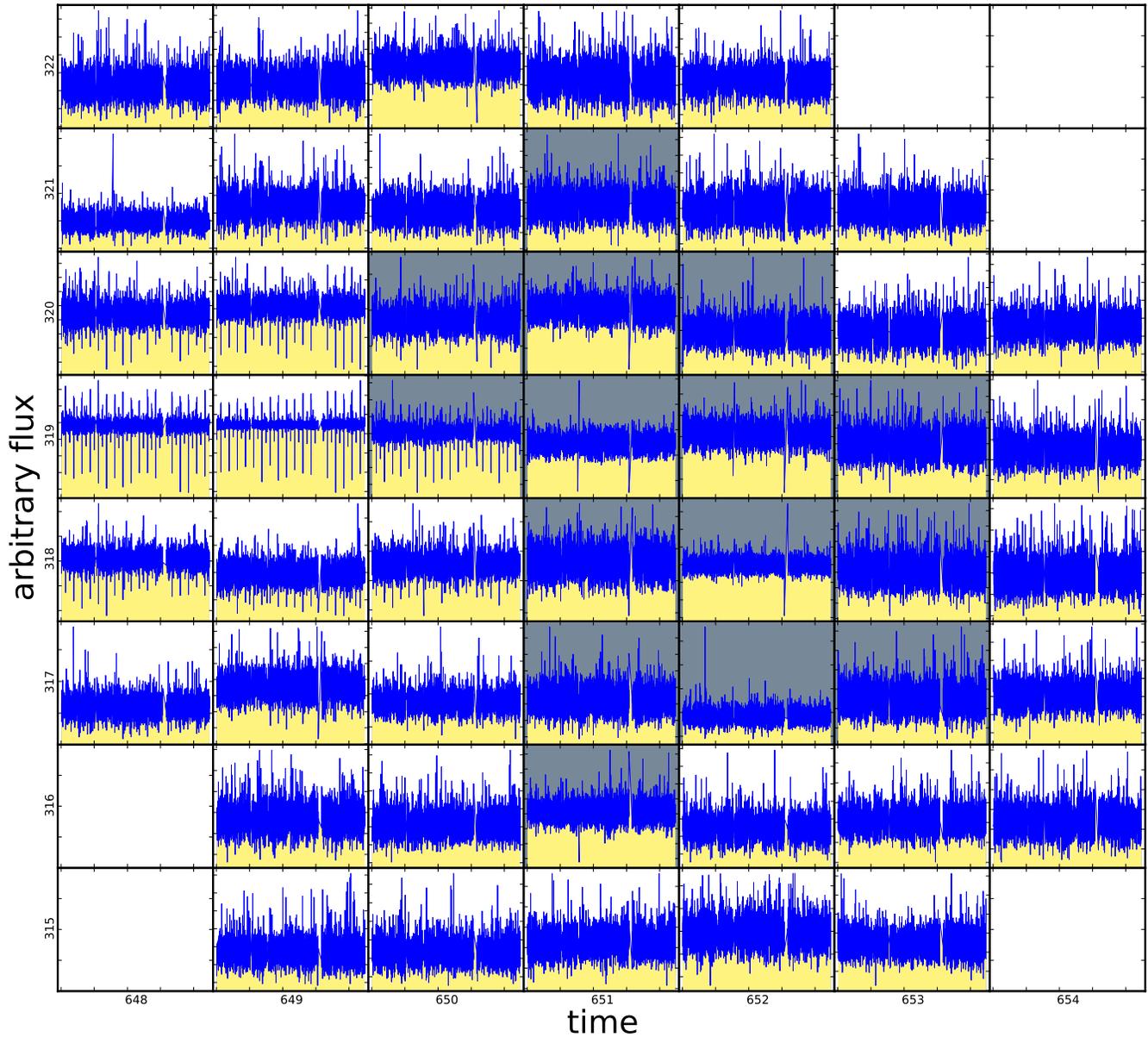}
\caption{Calibrated time-series photometry for every pixel within the mask surrounding the quarter 3 target KIC 2449074. Unlike Figure \ref{q2-tpf}, each light curve is plotted on an independent flux scale in order to identify if any background sources with respect to the brighter target appears. Source confusion with a background binary within the optimal aperture (gray pixels) is evident in pixel x=650, y=319.  The source of the regular dips found in the PDCSAP light curve of Figure \ref{lc-transits} is not the target that the mask was designed for. The source of the dips is a faint, background binary star on the left-hand side of the mask.}
\label{bg-eb-pix-zoom}
\end{figure}

This object shows regular, low amplitude dips in brightness every 4.9 days that, at face value, are suggestive of a planetary transit of the target star. Figure \ref{bg-eb-pix-zoom} shows a calibrated flux time series of each target mask pixel collected over Q3.  This figure was produced with the PyKE task \emph{keppixseries}:

\begin{quote}
{\tt keppixseries infile=kplr002449074-2009350155506\_lpd-targ.fits \\outfile=keppixseries.fits plotfile=keppixseries.png plottype=local filter=n}
\end{quote}

Figure \ref{bg-eb-pix-zoom} reveals unambiguously that the target star is not the source of the ''transit'' features. A background eclipsing binary star is situated 10 arcsec from the target star (2.5 pixels to the left of KIC 2449074 on the figure) and is leaking into the optimal aperture. By extracting the light curve manually using different pixels, one can either reduce the contaminating flux from the eclipsing binary or, alternatively, extract flux from the eclipsing binary. New mask files are created interactively using the \emph{kepmask} tool:

\begin{quote}
{\tt kepmask infile=kplr002449074-2009350155506\_lpd-targ.fits maskfile=mask1.txt plotfile=kepmask.png tabrow=2177 iscale=linear cmap=bone}

{\tt kepmask infile=kplr002449074-2009350155506\_lpd-targ.fits maskfile=mask2.txt plotfile=kepmask.png tabrow=2177 iscale=linear cmap=bone}
\end{quote}

The image associated with the 2,177th timestamp in the Target Pixel File is plotted over a linear intensity scale using the {\it ’bone’ color} lookup table.  Users of the {\it kepmask} tool define a new aperture interactively by moving the mouse over a pixel. One press of the 'X' keyboard key selects a pixel for inclusion within the new aperture, a second press deselects the pixel. The aperture is stored by clicking the 'DUMP' button on the interactive GUI. The task {\it kepextract} is employed to sum the pixels without weights within the newly defined aperture: 

In each case a new mask was defined interactively using the method described in Appendix B. Figure \ref{mask-targ} plots the mask stored in the file mask1.txt, while in Figure \ref{mask-eb} plots the mask stored in file mask2.txt. The commands used to extract new SAP light curve from the TPF are:

\begin{quote}
{\tt kepextract infile=kplr002449074-2009350155506\_lpd-targ.fits maskfile=mask1.txt outfile=kepextract1.fits}

{\tt kepextract infile=kplr002449074-2009350155506\_lpd-targ.fits maskfile=mask2.txt outfile=kepextract2.fits}
\end{quote}

The resulting SAP light curves of a less-contaminated target star and the background eclipsing binary star can be found in Figures \ref{lc-targ} and \ref{lc-eb}. Both were constructed using the {\it kepdraw} task:

\begin{quote}
{\tt kepdraw infile=kepextract1.fits outfile=kepextract1.png datacol=SAP\_FLUX ploterr=n errcol=SAP\_FLUX\_ERR quality=y }

{\tt kepdraw infile=kepextract2.fits outfile=kepextract2.png datacol=SAP\_FLUX ploterr=n errcol=SAP\_FLUX\_ERR quality=y }
\end{quote}

\section{Example 2: Mitigating Spacecraft Systematics by Re-extracting Target Pixels}

Preceding sections of this paper have provided qualitative motivation for replacing archived light curves with photometry re-extracted from the Target Pixel Files. From the TPFs, a customized light curve can be extracted from a new aperture containing any or all of the pixels in the target mask using a combination of the PyKE tasks {\it kepmask} (to define the new optimal aperture) and {\it kepextract} (to construct simple aperture photometry across the newly-defined aperture). In this example we consider the quarter 2 data for KIC 8703536, applying a custom aperture to extract a light curve from the pixel images. This target provides a conspicuous example because the source is spatially extended. The archived mask and aperture were constructed unaware of this fact and we can predict that the archived light curve flux is undercaptured. Furthermore, the KIC indicates that the target lies close to several fainter stellar sources that might be contaminants to the archived light curve. 

\begin{figure}
\includegraphics[scale=0.6,angle=90]{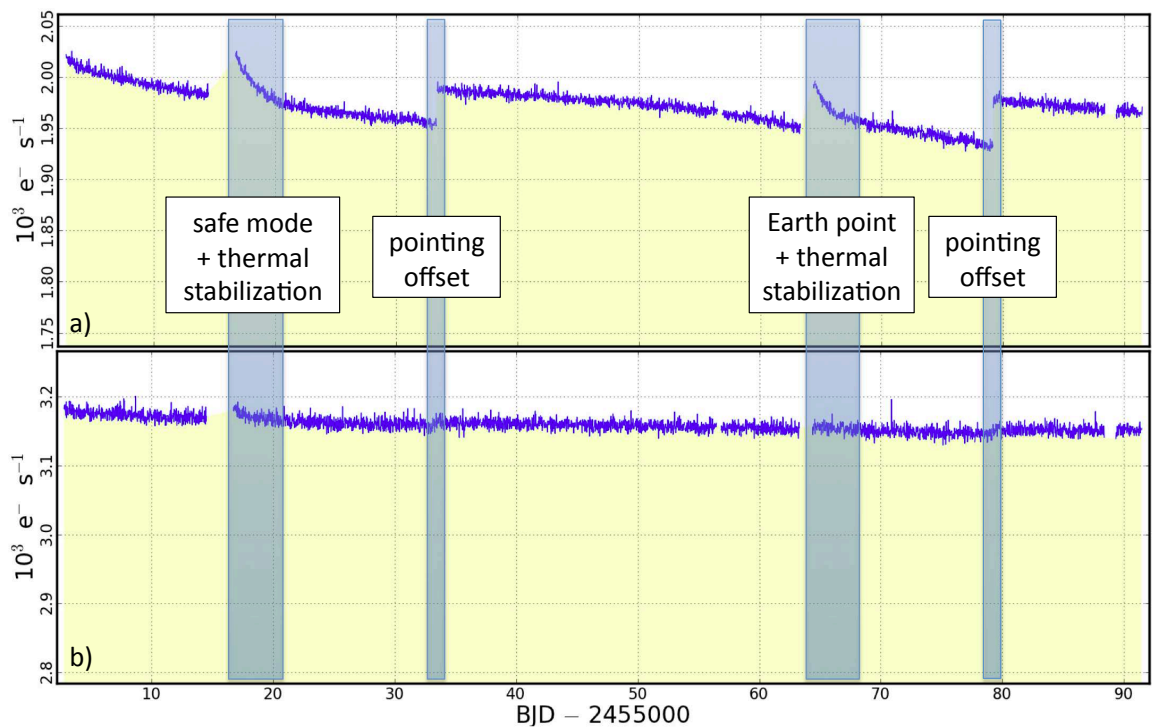}
\caption{a) The archived quarter 2 SAP light curve for the Kepler target KIC 8703536. Blue regions identify specific events adding conspicuous systematic noise to the light curve, as labelled within the white boxes. b) A version of the light curve, mitigated for systematics by a different choice of optimal aperture pixels.  A recipe for creating this new light curve from the archived Target Pixel File is provided in Appendix B.}
\label{kepmask}
\end{figure}

The archived SAP light curve for KIC 8703536 is displayed in the top panel of Figure \ref{kepmask}. This plot can be recreated using the PyKE task {\it kepdraw}: 

\begin{quote}
{\tt kepdraw infile=kplr008703536-2009259160929\_llc.fits outfile=sap.png datacol=SAP\_FLUX ploterr=n errcol=SAP\_FLUX\_ERR quality=y }
\end{quote}

The target is the Seyfert 2 galaxy 2MASX J19471938+4449425 and it is expected to be relatively quiet in the Kepler bandpass at high frequencies. Nevertheless, the SAP light curve displays four features that we can identify as systematic in nature because they coincide with spacecraft events and occur within all Kepler target light curves at the same time to a lesser or greater degree. These occur after a spacecraft safe mode, a pointing to Earth for data transfer, and two spacecraft attitude tweaks. Each systematic event can be identified by referencing the data quality flags supplied within the light curve file and TPF.

\begin{figure}
\includegraphics[scale=0.6,angle=0]{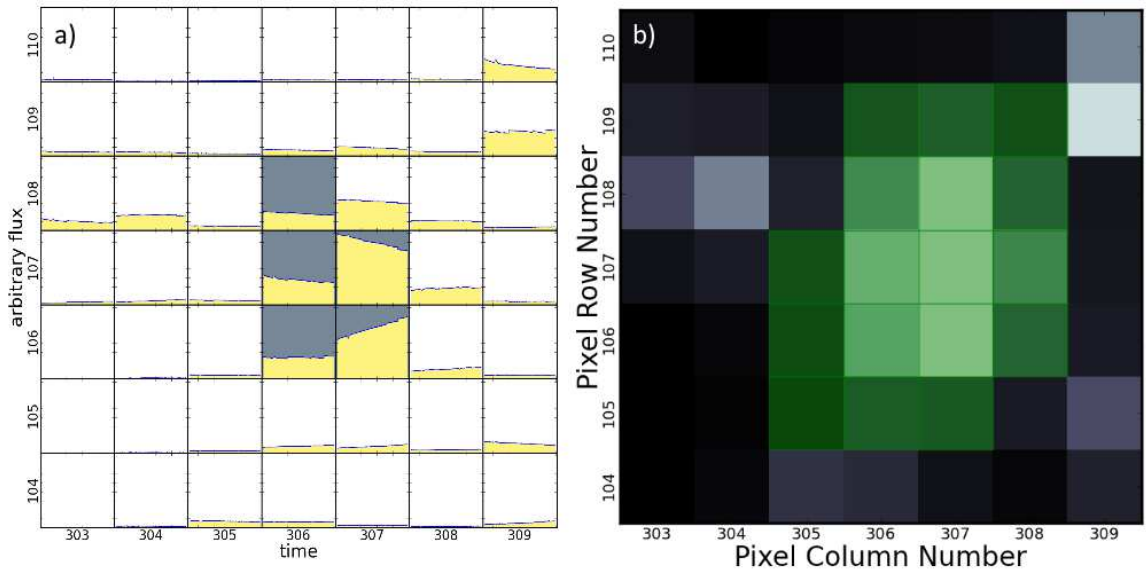}
\caption{a) Individual light curves from the pixels within the mask collected for the quarter 2 target KIC 8703536. The optimal aperture used to construct the archived light curve of this target (Figure \ref{kepmask}a) is defined by the grey pixels.  This figure was produced with PyKE task {\it keppixseries}, similar to Figure \ref{q2-tpf}. b) A manually-constructed aperture is defined by the green pixels, overlaid upon one specific postage stamp image contained within the archived Target Pixel File. The new aperture yields the light curve provided in Figure \ref{kepmask}b.  This figure was produced with PyKE task {\it kepmask}, which allow the user to select pixels to construct a new optimal aperture.  The selected pixels appear green over the postage stamp image of the target.}
\label{kepextract}
\end{figure}

The photometric aperture that yielded the light curve in Figure \ref{kepmask}a and the individual pixel photometry are provided in Figure \ref{kepextract}. This image can be replicated using the PyKE task {\it keppixseries}:

\begin{quote}
{\tt keppixseries infile=kplr008703536-2009259160929\_lpd-targ.fits \\outfile=keppixseries.fits plotfile=keppixseries.png plottype=global filter=n}
\end{quote}

The file kplr008703536-2009259160929\_lpd-targ.fits is the archived Target Pixel File coupled to the archived light curve. Two new files will be created -- keppixseries.fits contains the tabulated data for each individual pixel light curve, and keppixseries.png contains the requested plot. The parameter {\it plot type=global} requests that all light curves are plotted on the same photometric scale and the data plotted are not bandpass filtered to reduce the low frequency effects of differential velocity aberration. 

Our task is to reduce the systematic artifacts by extracting a new light curve from a more strategic choice of pixel aperture. The righthand panel of Figure \ref{kepextract} contains the pixel image from one specific timestamp in the Quarter 2 Target Pixel File of KIC 8703536. The green, transparent pixels represent a new custom photometric aperture defined using the PyKE task {\it kepmask}. An interactive image is called with the command: 

\begin{quote}
{\tt kepmask infile=kplr008703536-2009259160929\_lpd-targ.fits maskfile=mask.txt plotfile=kepmask.png tabrow=2177 iscale=linear cmap=bone}
\end{quote}

As described in Appendix A, in this example, the new aperture is stored in a filed called mask.txt, and the task {\it kepextract} is called for this target pixel file.

\begin{quote}
{\tt kepextract infile=kplr008703536-2009259160929\_lpd-targ.fits maskfile=mask.txt outfile=kepextract.fits}
\end{quote}

The light curve in the lower panel of Figure \ref{kepmask} is the target data re-extracted from the new aperture and plot, as before with {\it kepdraw}:

\begin{quote}
{\tt kepdraw infile=kepextract.fits outfile=kepextract.png datacol=SAP\_FLUX ploterr=n errcol=SAP\_FLUX\_ERR quality=y }
\end{quote}

While there remains some high frequency structure associated with the safe mode and potentially some residual low frequency noise related to differential velocity aberration, systematic artifacts are much reduced with the new aperture. Either the extended target is still not fully captured or we have introduced new systematic noise with the inclusion of new source contaminants. Optimizing the light curve further with additional aperture iterations is left as an exercise for the reader.

\section{Example 3: Systematic artifact removal using the cotrending basis vectors}

\begin{figure}
\includegraphics[width=\textwidth]{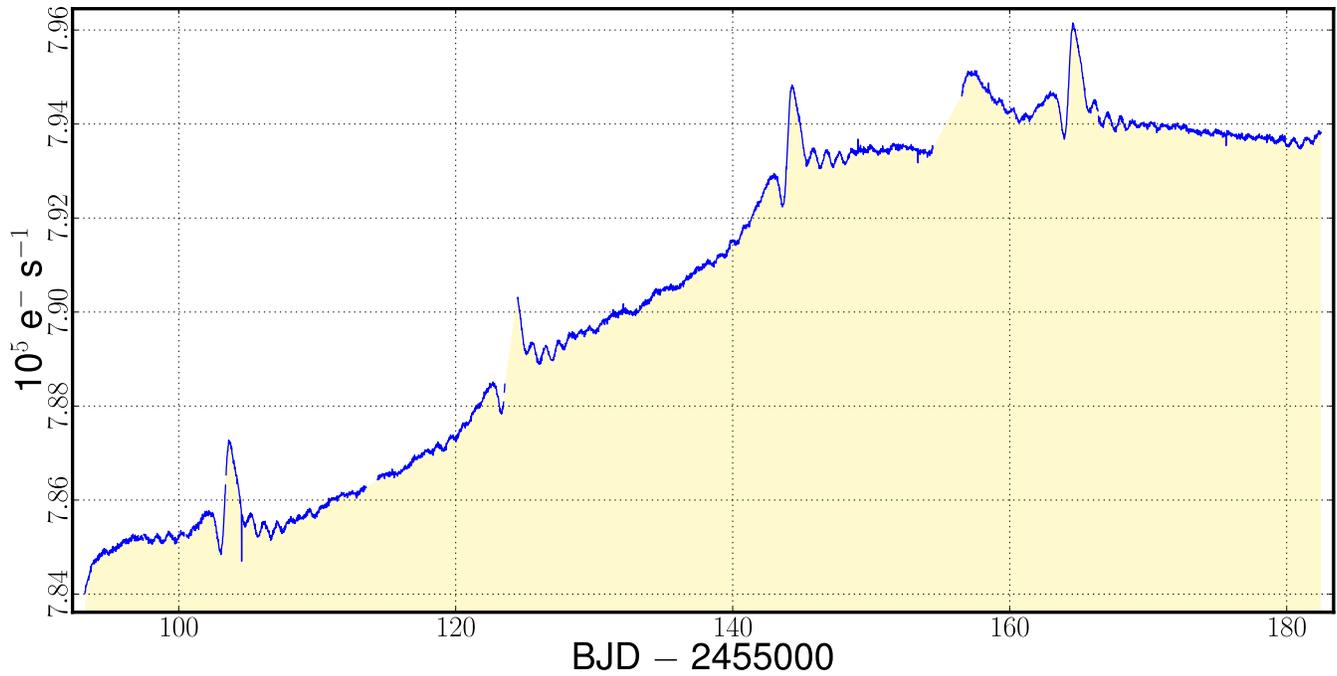}
\caption{The quarter 3 long cadence SAP light curve of KIC 3749404. The long-term trend of increasing flux, 1\% in amplitude,  is most-likely caused by differential velocity aberration. A 5-d interval of thermal settling after an Earth point at BJD 2,455,156.5 stands out as a likely systematic feature over the astrophysical signal. Cotrending Basis Vectors can be employed to remove or reduce all the undesirable systematic effects.}
\label{cbv-before}
\end{figure}

In this example we will reduce the systematic trends present in the quarter 3 SAP time-series of an eccentric binary star. As we proceed through the steps, note that each improvement requires a subjective decision based upon both foreknowledge of events recorded in the Kepler data quality flags and physical insight of the target in question. The SAP photometry of this target is plotted against barycenter-corrected time in Figure \ref{cbv-before}. This plot was made using the PyKE tool {\it kepdraw}: 

\begin{quote}
{\tt  kepdraw infile=kplr003749404-2009350155506\_llc.fits \\outfile=kplr003749404-2009350155506\_llc.png datacol=SAP\_FLUX ploterr=n errcol=SAP\_FLUX\_ERR quality=n}
\end{quote}

We fit the first two CBVs to the data using the {\it kepcotrend} task invocation:
\begin{quote}
{\tt kepcotrend infile=kplr003749404-2009350155506\_llc.fits \\outfile=kplr003749404-2009350155506\_cbv.fits \\cbvfile=kplr2009350155506-q03-d04\_lcbv.fits vectors='1 2' method=llsq iterate=n }
\end{quote}

\begin{figure}
\includegraphics[width=\textwidth]{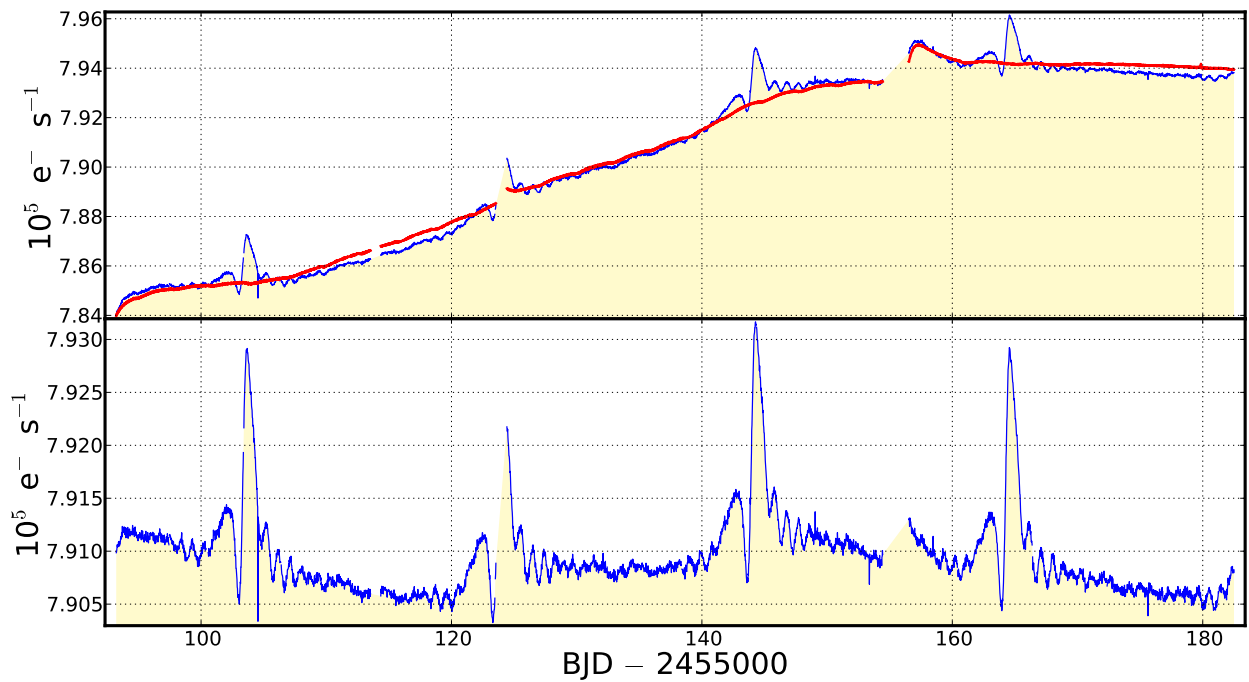}
\caption{A two-CBV fit to the archived quarter 3 SAP light curve of KIC 3749404. The upper panel of the plot shows the original SAP light curve in blue and the best linear least-squares fit of the two basis vectors in red. The lower plot contains the result of subtracting the basis vectors fit from the original light curve. While systematic effects have been minimized, some remain. For example, the poor basis vector fit to the data between the first and second maxima around BJD 2,455,115, shows that the systematics were not completely mitigated. }
\label{twofit}
\end{figure}

\begin{figure}
\includegraphics[width=\textwidth]{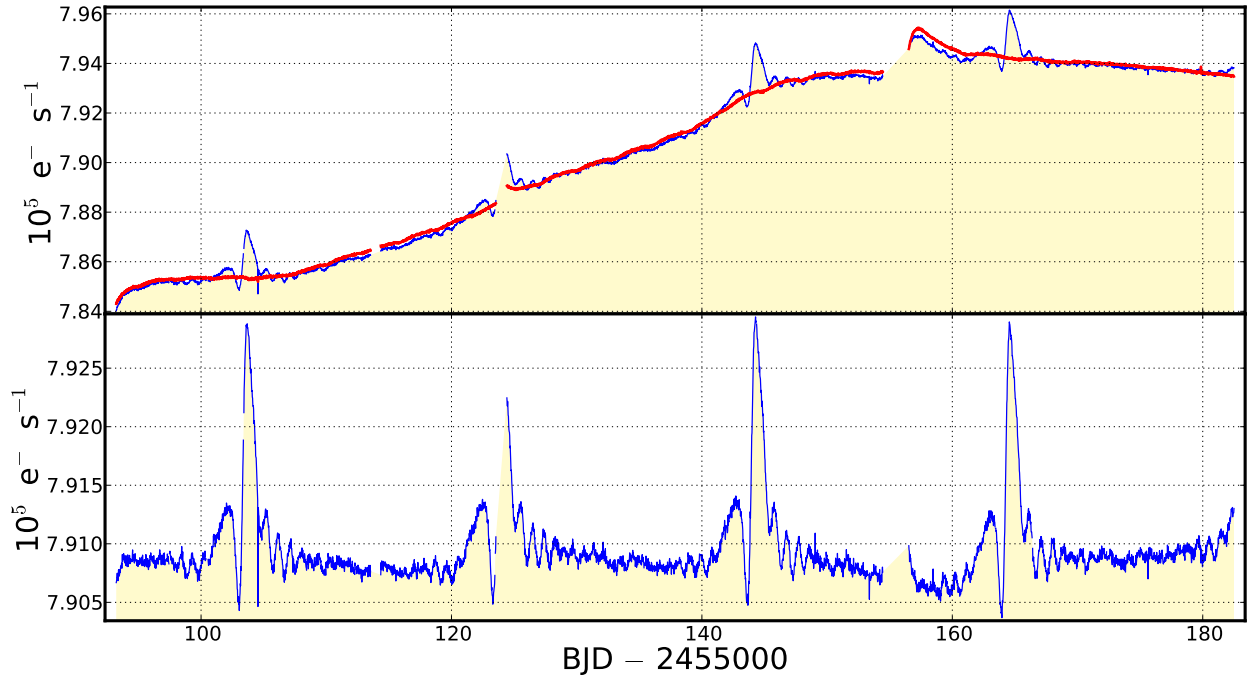}
\caption{As for Figure \ref{twofit}. This time, we fit five CBVs to the quarter 3 SAP light curve of KIC 3749404. The systematics now appear to be much reduced but there are still some effects in the second half of the quarter that can be mitigated further (e.g. the poor fit to the thermal settling event around BJD 2,455,156.5.)}
\label{fivefit}
\end{figure}

The {\it llsq} method requires {\it kepcotrend} to perform a linear least-squares fit and subtraction of the basis vectors upon the SAP data. No sigma clipping iterations are performed during the fit. The quarter 3 CBV file, kplr2009350155506-q03-d04\_lcbv.fits, can be downloaded from the Kepler archive at MAST. The full content of the input light curve file is copied to the output file and a new column called CBVSAP\_FLUX is appended to the FITS table containing the best-fit, CBV-subtracted light curve. The result is shown in Figure \ref{twofit} and yields an improvement over the photometric quality of the SAP light curve. The long-term trend has been greatly reduced, but there are still higher-frequency features that are most likely systematic, and the fit can be improved further. In particular, we would like to remove the broad dip that has been introduced between the first and second brightening events in the time-series.  We will strive to obtain a correction where the heights of each event are identical. Performing another fit using five basis vectors with the following command yields the result shown in Figure \ref{fivefit}:

\begin{quote}
{\tt kepcotrend infile=kplr003749404-2009350155506\_llc.fits \\outfile=kplr003749404-2009350155506\_cbv.fits \\cbvfile=kplr2009350155506-q03-d04\_lcbv.fits vectors='1 2 3 4 5' method=llsq iterate=n}
\end{quote}

\begin{figure}
\includegraphics[width=\textwidth]{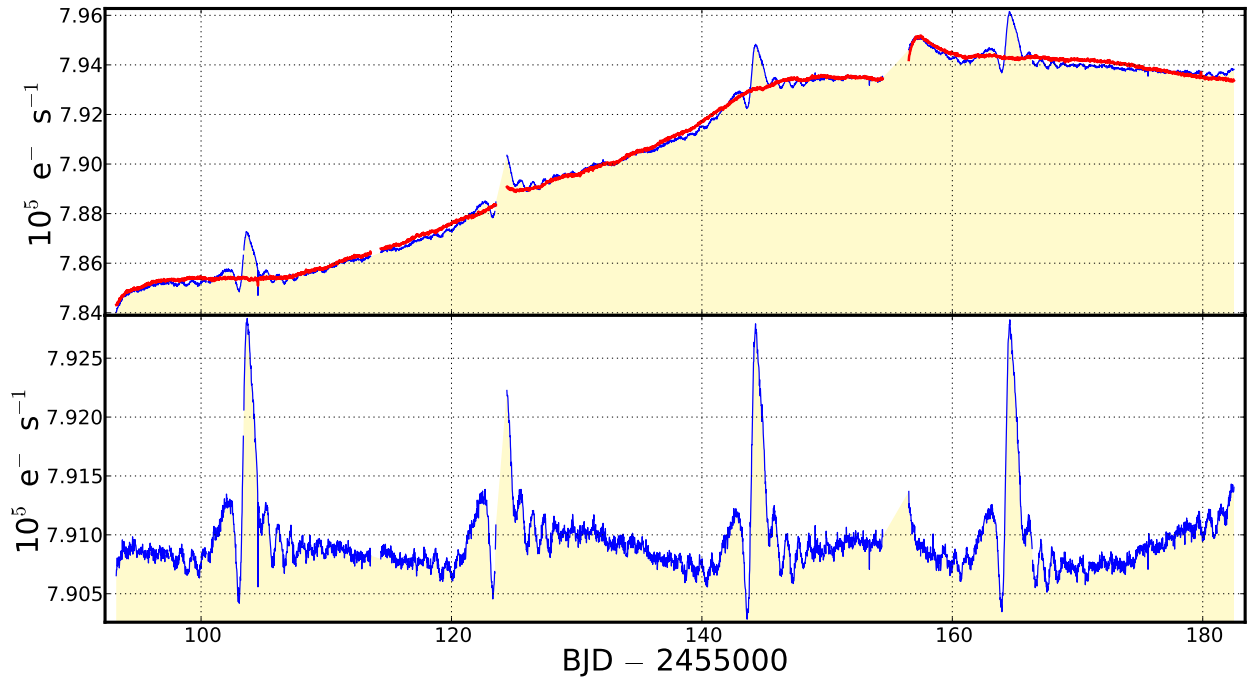}
\caption{As for Figures \ref{twofit} and \ref{fivefit}. This fit to the quarter 3 SAP light curve of KIC 3749404 contains eight basis vectors and appears to be performing less well than a five basis vector fit around the last brightening event.}
\label{eightfit}
\end{figure}

This new result is a qualitative improvement compared to the two-CBV fit, but the solution is still not optimal. We would like to improve the fit to the thermal event at BJD 2,455,156.5 and also further flatten the structure occurring after BJD 2,455,145. We fit the SAP data again, this time using eight basis vectors. The plot is shown in Figure \ref{eightfit} but the result appears to be less optimal than the 5 basis vector fit. Anomalous structure has very likely been added to the time series by the CBVs. One possible reason for the less than optimal solution is that eight basis vectors are over-fitting the periodic brightenings and adding new systematic noise to the intervals between them. To test this hypothesis we masked out light curve segments containing the large amplitude brightenings and fit the CBV to the remaining data. We employed the PyKE task {\it keprange} to define the masked regions in the time series:

\begin{quote}
{\tt keprange infile=kplr003749404-2009350155506\_llc.fits outfile=keprange.txt column=SAP\_FLUX}
\end{quote}

\begin{figure}
\includegraphics[width=\textwidth]{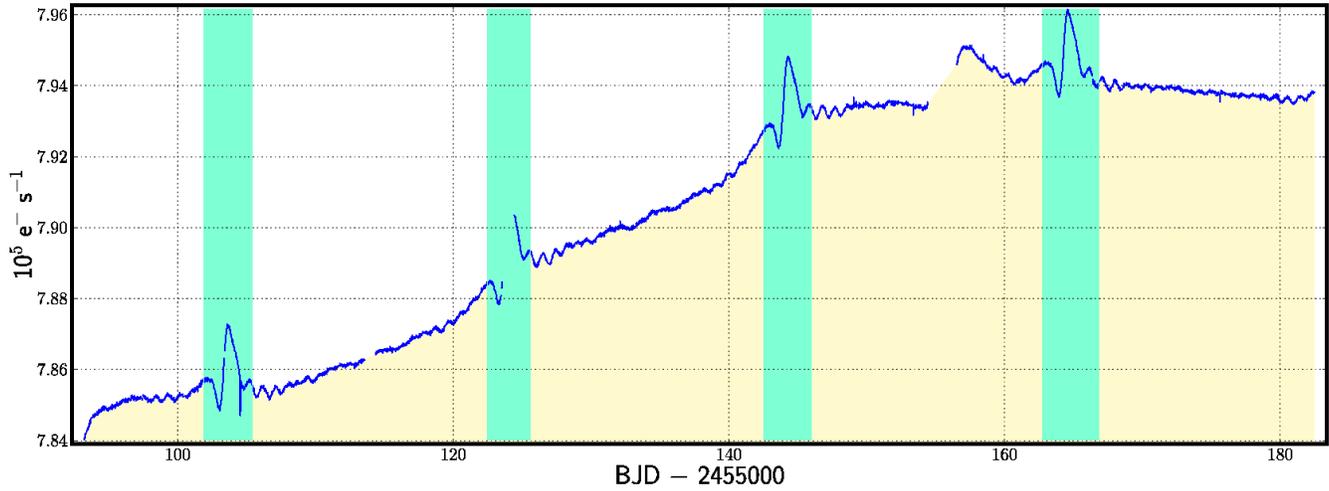}
\caption{The flux level in the green highlighted regions is not typical of the rest of the light curve and posed problems for the eight basis vector least-square fit. In order to improve the CBV fit to the SAP light curve we ignored the regions highlighted in green during fit minimization. This figure contains the interactive environment of the {\it keprange} tool, developed to define discrete regions of time-series data.}
\label{keprange}
\end{figure}

This will plot the SAP\_FLUX column data within the light curve file over time. Ranges in time can be defined by selecting start and stop times with the mouse and 'X' keyboard key. We masked four ranges in this example, as illustrated in Figure \ref{keprange}, and these ranges will be saved to a text file after clicking the `SAVE' button on the interactive GUI.  We performed the eight basis vector fit one last time, excluding from the fit the regions defined in Figure \ref{keprange}, again using the {\it kepcotrend task}: 

\begin{quote}
{\tt kepcotrend infile=kplr003749404-2009350155506\_llc.fits \\outfile=kplr003749404-2009350155506\_cbv.fits \\cbvfile=kplr2009350155506-q03-d04 lcbv.fits vectors='1 2 3 4 5 6 7 8' method=llsq iterate=n maskfile=keprange.txt}
\end{quote}

\begin{figure}
\includegraphics[width=\textwidth]{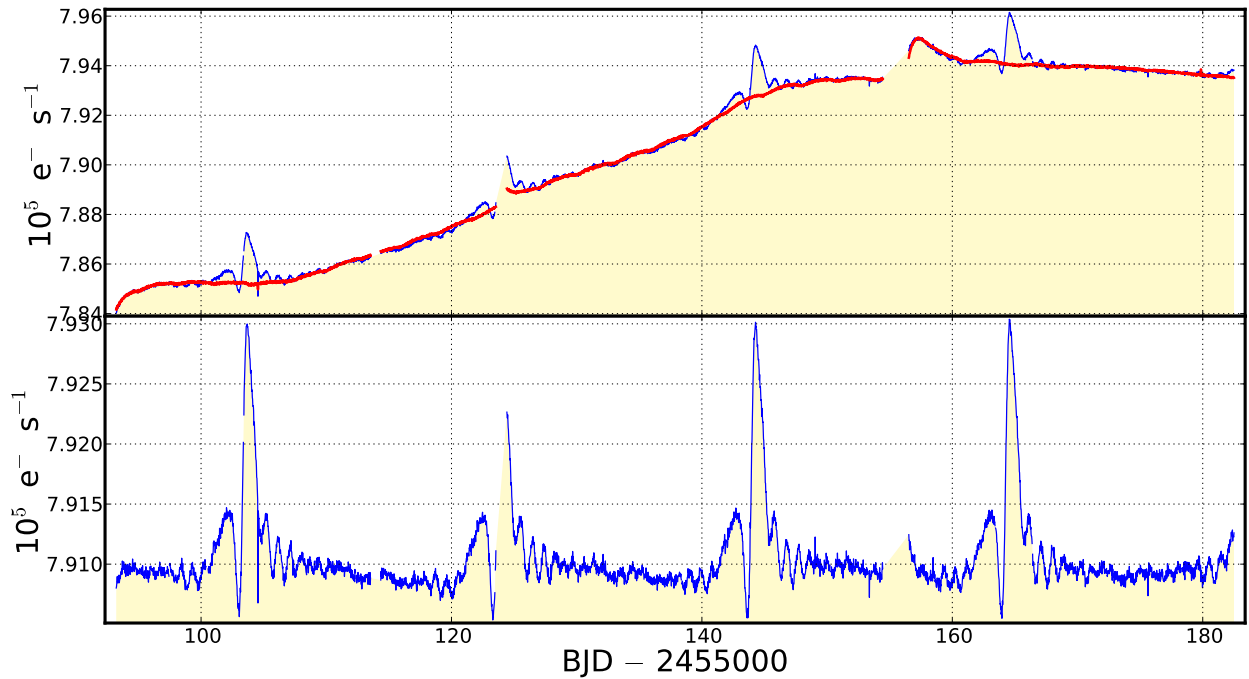}
\caption{The final iteration of the CBV fit to the quarter 3 SAP light curve of KIC 3749404. We used the PyKE tool {\it kepcotrend} and fit eight basis vectors to the light curve. We did not fit the regions of the light curve highlighted in Figure \ref{keprange}.}
\label{kepcotrend}
\end{figure}

In Figure \ref{kepcotrend} we see the final version of the light curve. Systematic effects still remain, e.g. the thermal settling event is not totally removed but, subjectively, the data are much improved over the time series. The quality of the corrected light curve here was considered adequate for the scientific analysis presented by \citet{Thompson:2012}.

\bibliography{kepbib2}

\clearpage


\begin{figure}
\includegraphics[scale=0.4,angle=0]{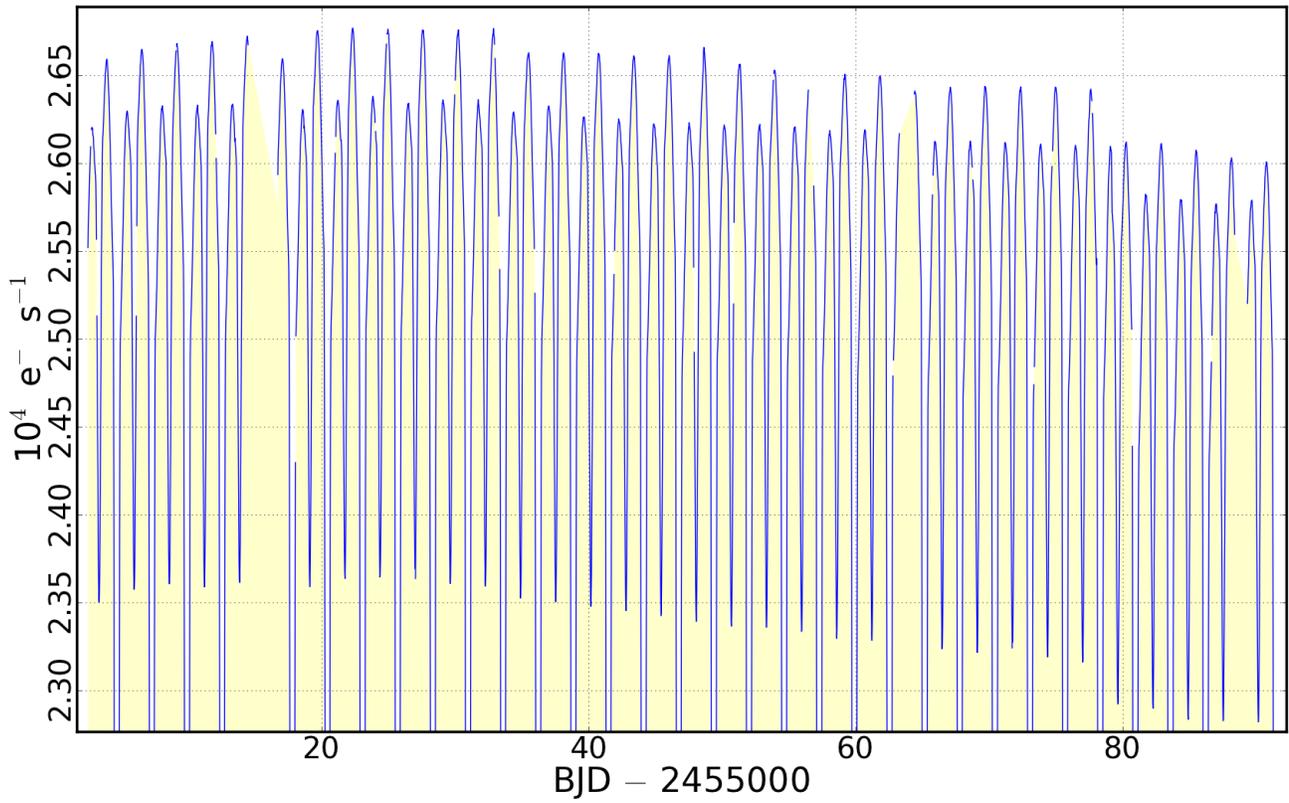}
\caption{Quarter 2 long cadence SAP light curve of the eclipsing binary star V1950 Cyg (KIC 12164751; \citet{Horne:2008}), produced by the PyKE tool {\it kepdraw}. The most-prominent systematic effect in this light curve is the long-term decay in flux from the target which falls by 3\% over the duration of the quarter. This drop is a consequence of differential velocity aberration pushing the under-captured target position across the fixed pixel aperture over time.}
\label{q2-sap}
\end{figure}

\begin{figure}
\includegraphics[width=\textwidth]{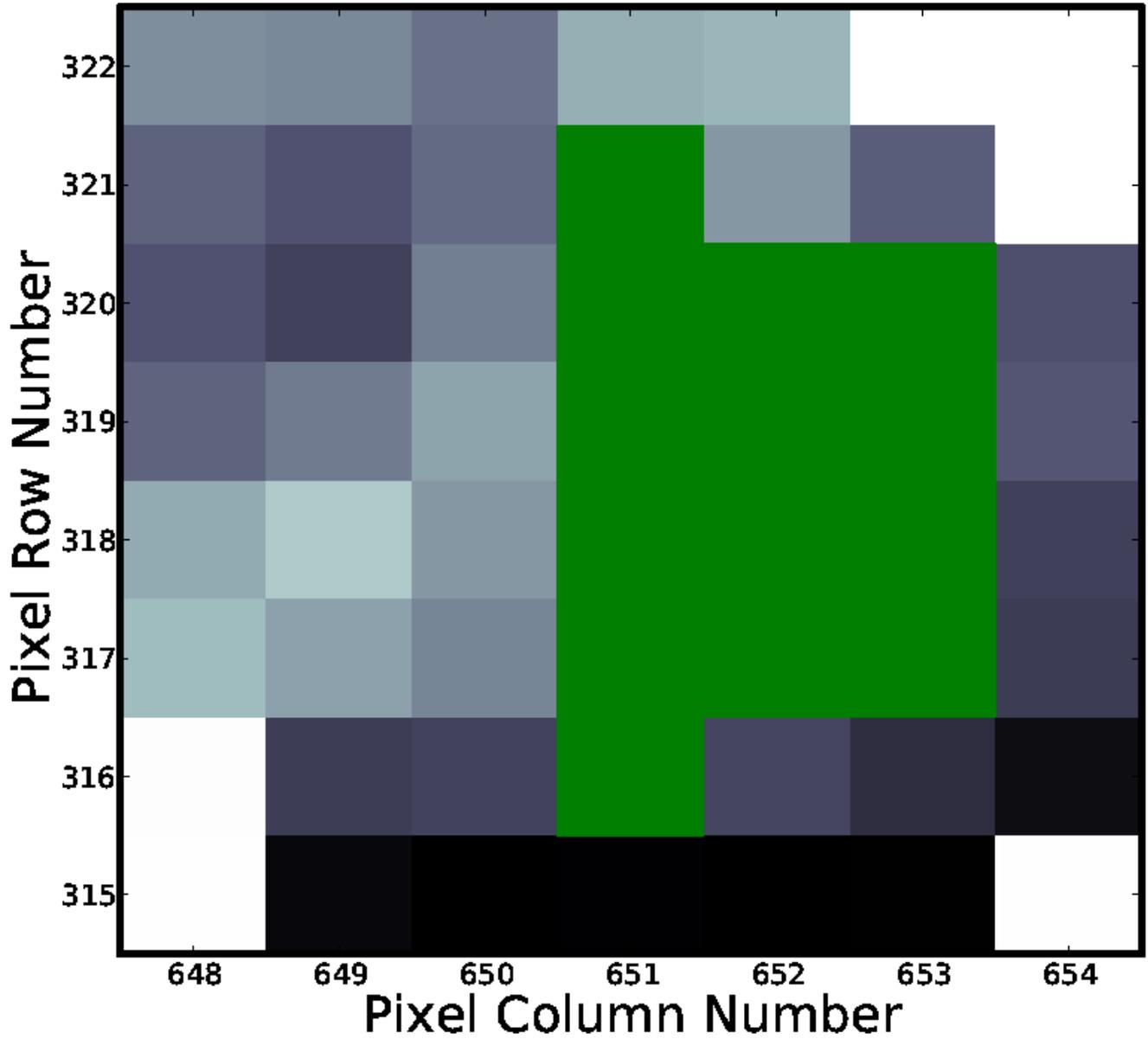}
\caption{This pixel map image, created with the PyKE tool {\it kepmask}, is a typical flux image within the quarter 3 pixel mask of KIC 2449074. The green region is a manually-defined photometric aperture that maximizes the signal from the target star. The new selection of pixels, as indicated in green, minimizes the contamination from a background eclipsing binary within the target mask. When summed, the pixels within the new optimal aperture produce the light curve in Figure \ref{lc-targ}.}
\label{mask-targ}
\end{figure}

\begin{figure}
\includegraphics[width=\textwidth]{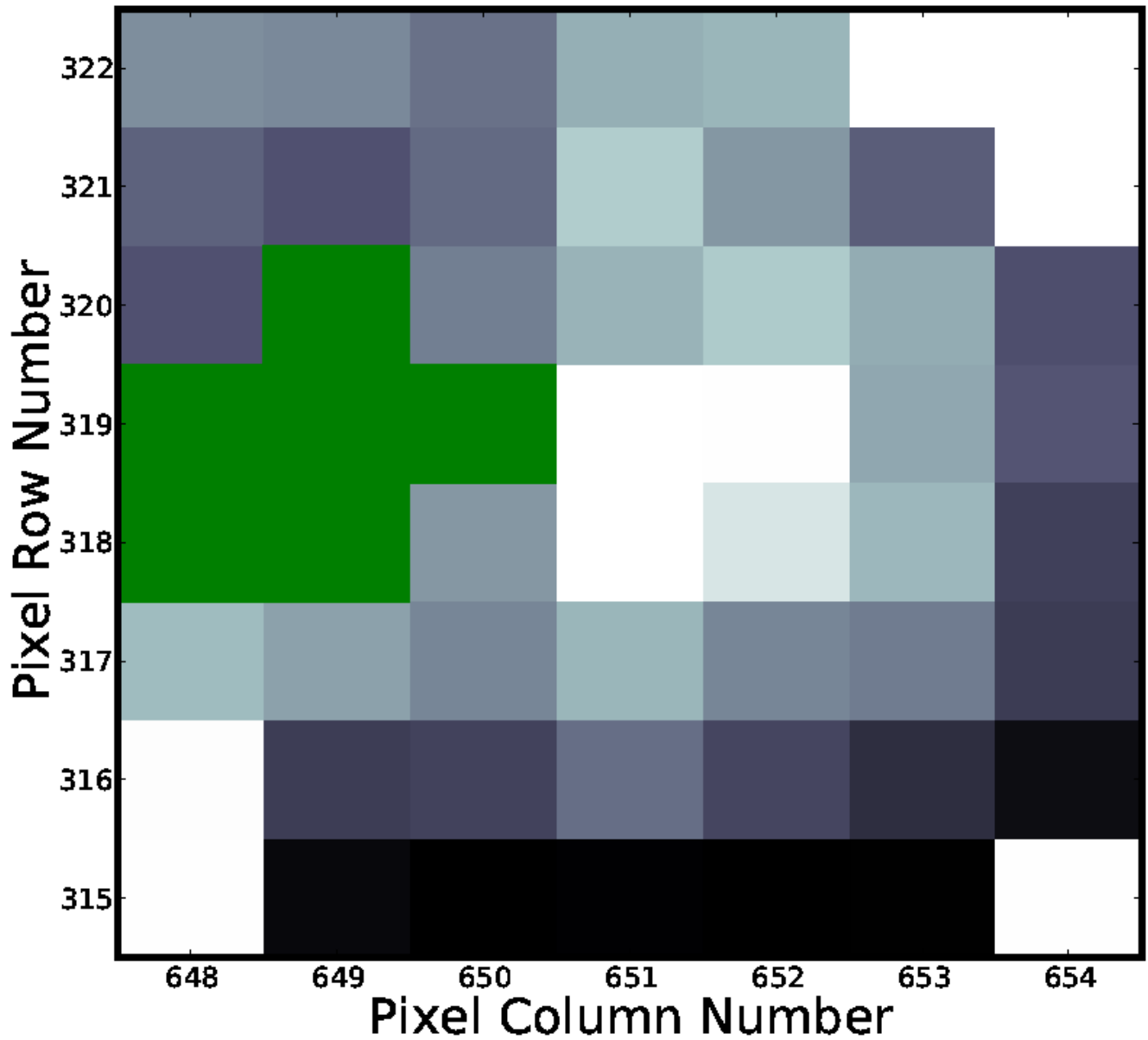}
\caption{As for Figure \ref{mask-targ}, except the green region is a manually-defined optimal aperture that maximizes the signal from the background binary star within the mask. The selected green pixels minimize the contamination from the target star. When summed, the pixels within the new optimal aperture produce the light curve in Figure \ref{lc-eb}.}
\label{mask-eb}
\end{figure}

\begin{figure}
\includegraphics[width=\textwidth]{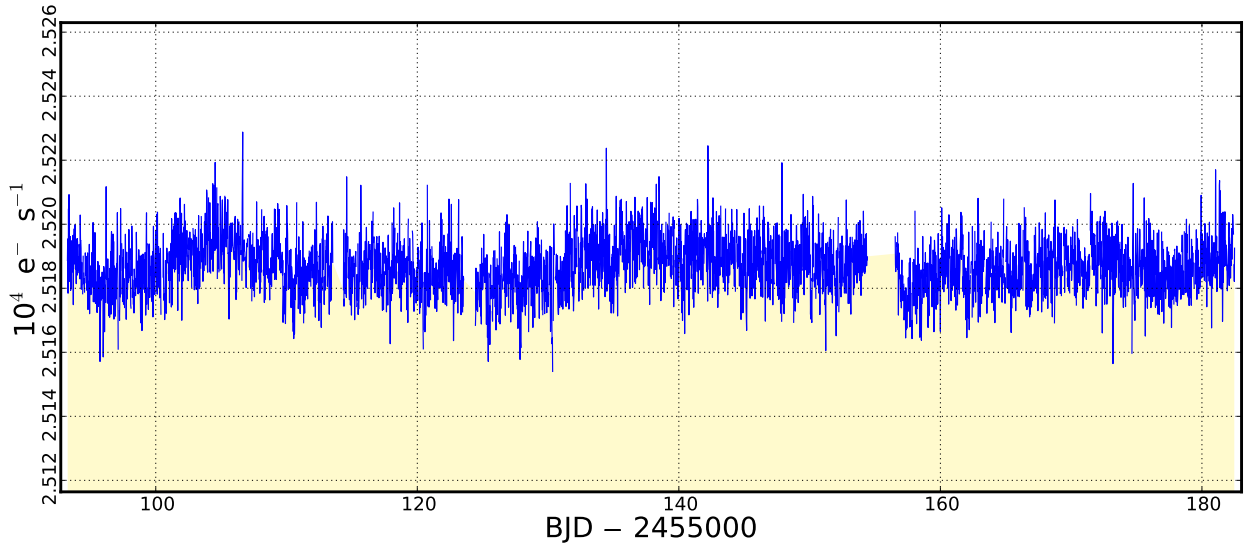}
\caption{The quarter 3 SAP light curve for  KIC 2449074, constructed manually by summing the pixels within the optimal aperture defined in Figure \ref{mask-targ}. The new user defined aperture minimizes the contamination from the nearby background binary, evident in Figure \ref{bg-eb-pix-zoom}.}
\label{lc-targ}
\end{figure}

\begin{figure}
\includegraphics[width=\textwidth]{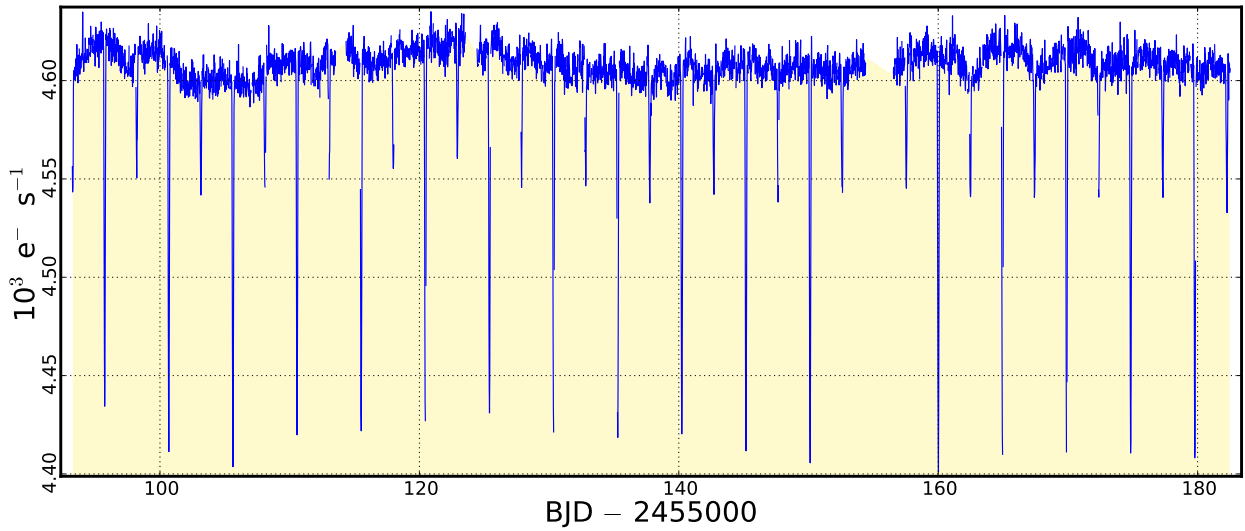}
\caption{The quarter 3 SAP light curve for the background binary within the target mask of KIC 2449074, constructed by summing the pixels within the optimal aperture defined in Figure \ref{mask-eb}.}
\label{lc-eb}
\end{figure}


\end{document}